\documentclass[11pt]{article}
\pdfoutput=1

\usepackage{jcappub}

\usepackage{amsfonts} 
\usepackage{amsmath} 

\usepackage{graphicx} 
\usepackage{float}
\usepackage[margin=1in]{geometry}
\usepackage{float}
\usepackage{tabularx}
\usepackage{array}
\usepackage{epstopdf}

\usepackage{caption}
\usepackage{subfig}

\DeclareMathOperator{\erf}{erf}

\title{Vacuum Selection on Axionic Landscapes}

\author{Gaoyuan Wang,}
\author{Thorsten Battefeld}
\affiliation{Institute for Astrophysics, University of Goettingen, Friedrich Hund Platz 1, D-37077 Goettingen, Germany}

\emailAdd{gaoyuan.wang@stud.uni-goettingen.de,\\
tbattefe@astro.physik.uni-goettingen.de}

\abstract{We compute the distribution of minima that are reached dynamically on multi-field axionic landscapes, both numerically and analytically. Such landscapes are well suited for inflationary model building due to the presence of shift symmetries and possible alignment effects (the KNP mechanism). 

The resulting distribution of dynamically reached minima differs considerably from the naive expectation based on counting all vacua. These differences are more pronounced in the presence of many fields due to dynamical selection effects: while low lying minima are preferred as fields roll down the potential, trajectories are also more likely to get trapped by one of the many nearby minima. We show that common analytic arguments based on random matrix theory in the large $D$-limit to estimate the distribution of minima are insufficient for quantitative arguments pertaining to the dynamically reached ones. This discrepancy is not restricted to axionic potentials.  We provide an empirical expression for the expectation value of such dynamically reached minimas' height and argue that the cosmological constant problem is not alleviated in the absence of anthropic arguments. We further comment on the likelihood of inflation on axionic landscapes in the large $D$-limit. }

\arxivnumber{xxxx.xxxx}

\begin{document}
\maketitle
\flushbottom

\section{Introduction}
\label{sec:intro}

The use of axions for inflationary model building has a long history, ranging from initial attempts to use the QCD axion and  natural inflation \cite{Freese:1990rb} to multi-axion potentials in string theory as in N-flation \cite{Dimopoulos:2005ac} or monodromy inflation
 \cite{McAllister:2008hb} among many other proposals, see \cite{Baumann:2014nda,Marsh:2015xka} for recent reviews. Axions are popular candidates to drive inflation due to their (broken) shift symmetry, which protects the inflationary potential from corrections that would spoil inflation otherwise. It was soon realized that the QCD axion, as well as single field models in string theory, can not support large field inflation while remaining consistent with theoretical constraints and observations, see e.g.~the current lower bound in \cite{Planck:2013jfk} on the axion decay constant $f$, based on the Planck satellite mission that measured temperature fluctuations in the cosmic background radiation. This requirement is inconsistent with constraints in string theory \cite{Banks:2003sx,Svrcek:2006yi}, see also \cite{Rudelius:2014wla,Grimm:2014vva} and \cite{Baumann:2014nda}, so that attention shifted to multi-field models. Here, alignment effects, such as the Kim-Nilles-Peloso mechanism \cite{Kim:2004rp}, see also \cite{Peloso:2015dsa,Long:2014dta,Choi:2014rja}, enable the construction of large-field slow-roll inflationary models \cite{Bachlechner:2014gfa,Bachlechner:2015qja} that are consistent with observations  \cite{Planck:2013jfk} and appear to be under computational control \cite{Bachlechner:2014gfa,Bachlechner:2015qja,Kaloper:2015jcz}\footnote{ See however \cite{Rudelius:2014wla,delaFuente:2014aca,Montero:2015ofa,Brown:2015iha,Heidenreich:2015wga,Rudelius:2015xta,Brown:2015lia,Junghans:2015hba,Kappl:2015esy,Heidenreich:2015nta} for constraints on axionic inflation based on the weak gravity conjecture.}.

While inflation in such multi-field scenarios is certainly possible, it is not a priory clear how likely sufficiently long bursts of inflation are, or at which height the final vacuum after inflation sits. In the presence of many fields, whose masses and couplings constitute free parameters, we are essentially dealing with random potentials, see \cite{Tegmark:2004qd,Aazami:2005jf,Easther:2005zr,Frazer:2011tg,Battefeld:2012qx} for a small selection of relevant work. As a consequence, the first question is often addressed by sampling an ensemble of potentials and/or trajectories \cite{Aazami:2005jf,Frazer:2011tg} while invoking anthropic arguments \cite{Freivogel:2005vv} to focus only on those trajectories that support at least sixty e-folds of inflation. Distributions for observables can be computed in a similar manner, see \cite{Price:2015qqb,Frazer:2011br,McAllister:2012am,Dias:2012nf,RenauxPetel:2012ph,Westphal:2012up,Pedro:2013nda,Battefeld:2013xwa} for selected works (discrepancies between \cite{Dias:2012nf} and \cite{McAllister:2012am} are explained in \cite{McAllister:2012am}). One may also attempt to treat the functional form of the potential as random, see e.g. \cite{Jain:2015mma}.

In this article, we are interested in the final resting place, that is the resulting vacuum after dynamical evolution on an axionic potential, and only comment briefly on the feasibility of inflation. A common approach is to count all vacua in a given potential, see \cite{Aazami:2005jf,Chen:2011ac,Bachlechner:2012at,Higaki:2014mwa,Brodie:2015kza} for a few examples and \cite{Mehta:2015nva} for efficient methods.  Based on the resulting histogram, one attempts to draw conclusions about the likelihood of finding a vacuum at a particular height. If the distribution peaks at a height comparable to the observed value of the cosmological constant, one may not have to rely entirely on anthropic arguments \cite{Weinberg:1987dv,Bousso:2008bu}, see \cite{Polchinski:2006gy,Bousso:2012dk} for recent reviews \footnote{Note that even if the height were in the range of the observed vacuum energy, the cosmological constant (CC) problem would not be solved, since subsequent phase transition lead to additional large contributions of the CC. Nevertheless, achieving a small bare contribution to the CC after inflation appears to be a step in the right direction. An additional complication is the measure problem, see \cite{Freivogel:2011eg,Olum:2012bn,Schiffrin:2012zf} for a summary of proposed measures and reviews.}.

We show that \emph{simple counting arguments are misleading} because they do not take into account the  dynamical evolution of fields. As observed in \cite{Battefeld:2012qx} in a particularly simple toy model, low lying minima are usually more likely to be reached dynamically, particularly as the dimension of field space is increased.
In essence, what would be a minimum in a low dimensional potential is more often than not a saddle point in the higher dimensional case \footnote{If one holds one field fixed and finds a minimum for the other ones, that point is usually not a minimum if the fixed field is taken into consideration again. Naively, one might think the chance for a critical point to be a minimum to be  $2^{-D}$, but for many random potentials, this chance is actually suppressed by a factor of $e^{- \text{const} D^p}$ with $p=2$ as long as the Hessian can be approximated by a random Gaussian matrix (a known results in random matrix theory), see e.g.~\cite{Aazami:2005jf}.}, so that fields keep on rolling further and further down. This effect can be quantified numerically in concrete landscapes and understood analytically, at least qualitatively, based on random matrix theory (see i.e.~\cite{Mehta:1991,Rao:2005} for a textbook introduction to RMT) in the limit of many fields due to the feature of universality \cite{Deift:2006,Kuijlaars:2011,Bai,Soshnikov:2002,Schenker:2005}
\footnote{It should be noted that adding more structure to the scalar sector can render random matrix theory inapplicable, see e.g. the Type IIB flux compacitifcations examined in \cite{Brodie:2015kza}.}. However, a competing effect is the attraction to minima close to the trajectory: as the dimensionality is increased, more and more minima are close by, which can cause the trajectory to derail and get trapped early on. The interplay of these competing effect renders an analytic description challenging.

In this work, we consider a relatively simple axionic multi-field potential, Sec.~\ref{sec:thesetup}, similar to the one considered in \cite{Higaki:2014mwa,Higaki:2014pja}. We compute the distributions of dynamically reached vacua numerically in Sec.~{\ref{sec:numerical studies}}. To explain these distributions analytically, particularly the preference of low lying minima, we examine the statistical properties of the potential: we compute the probability distribution function (PDF) of the potential and its first derivative at a random point in Sec.~\ref{Sec:pdfpotential} and \ref{sec:PDFpotentialgradient}. In conjunction with an empirical ansatz (tested numerically) of the average potential difference to the next critical point $\overline{\Delta V}$, we estimate the number of critical points $n_{\mbox{c}}$ that are encountered if a given potential difference is traversed, see Sec.~\ref{subsec:difftocri}. We follow with the PDF of the Hessian in Sec.~\ref{subsec:Hessian}, using results of random matrix theory.  We pay particular attention to the many field limit, $D\gg 1$. Some technical aspects of the computation are relegated to the Appendix. 

Given $n_{\mbox{c}}$ and the PDF of the Hessian, we compute the expected distribution of minima in Sec.~\ref{subsec:whyrmtfails}. While the analytic result recovers some aspects properly, such as the presence of a peak and the shift of the peak to lower values as $D$ is increased, it is insufficient for a quantitative analysis. We identify the most likely causes for the observed discrepancies: the omission of dynamical effects as well as constraints and correlations between elements of the Hessian. 

Based on numerical studies, we are able to provide a simple empirical fitting function for the ratio of the average potential difference between critical points to the mean potential difference to the nearest minimum, $R(D)=dV / \overline{\Delta V}$, see Sec.~\ref{Sec:empirical}. In conjunction with our estimate for $\overline{\Delta V}$, we are able to compute the height of the most likely resting place (\ref{finalVmin}),  yielding
\begin{eqnarray}
V_{\text{min}}(D,n)\propto \left(1-\text{const.}\,\frac{\ln(15+D)}{\sqrt{n}}\right)\,,
\end{eqnarray}
where $n\gg D$ is the number of sources in the axionic potential. This result is in quantitative agreement with our numerical studies, but lacks a thorough theoretical justification, particularly the origin of the logarithmic dependence on $D$. 

We discuss implications for the cosmological constant problem in Sec.~\ref{disc:ccproblem}. We conclude that it is crucial to incorporate dynamical selection effects consistently if one is interested in the distribution of vacua that may be reached after inflation. Concretely, counting all vacua or relying on random matrix theory alone is insufficient for quantitative arguments.

We follow with a brief discussion of inflation on such landscapes in Sec.~\ref{inflation}: as the dimensionality of field space is increased, inflation becomes more likely. Due to additional numerical challenges in the large $D$ limit,  we postpone a detailed analysis to a forthcoming publication.  We conclude in Sec.~\ref{conclusion}. 

\section{Axionic Landscapes}
\label{sec:thesetup}
We consider multi-field axionic potentials of the form \cite{Higaki:2014mwa}
\begin{eqnarray}
V=\sum_{i=1}^{N_{\mbox{\tiny source}}}\Lambda_i^4\bigg(1-\cos\big(\sum_{j=1}^{N_{\mbox{\tiny axion}}}n_{ij}\frac{\phi_j}{f_j}+\theta_i\big)\bigg)+C\,,\label{defpotential}
\end{eqnarray}
where $N_{\mbox{\tiny axion}}$ is the number of axions, $N_{\mbox{\tiny source}}$ is the number of shift symmetry breaking sources from non-perturbative effects, $\Lambda_i$ sets the strength of each source term, $\phi_j$ denote the axions with decay constants $f_j$, $n_{ij}$ is a mixing matrix, $\theta_i$ the relative phases between different source terms and $C$ an overall additive constant affecting the value of the cosmological constant today.

Our notation is consistent with the one in \cite{Higaki:2014mwa}, in which the distribution of vacua was investigated via direct counting, taking into account constraints that enable the Kim-Nilles-Peloso (KNP) mechanism \cite{Kim:2004rp}, see also \cite{Berg:2009tg,Tye:2014tja,Ben-Dayan:2014zsa}. We pursue a similar goal, but would like to take into account dynamical effects: once a field is let to evolve on a given landscape, it rolls towards low lying minima, leading to a selection effect that is expected to be stronger the higher the dimensionality of field space is, see \cite{Battefeld:2012qx,Battefeld:2013xwa}. On the other hand, trajectories are more likely to get trapped early on as $N_{\text{axion}}$ increases, since minima close to the trajectory act as attractors. Throughout this article we set
\begin{eqnarray}
m_{\mbox{\tiny Pl}}\equiv \sqrt{\frac{1}{8\pi G}}\equiv 1\,.
\end{eqnarray}

\subsection{Model Parameters \label{sec:modelparameters}}
In line with \cite{Higaki:2014mwa}, we do not consider the most general potential, but impose several constraints on model parameters. Firstly, we set 
\begin{eqnarray}
C\equiv 0\,,
\end{eqnarray}
so that the lowest minimum is at $V=0$. Since it is unlikely that the axions  settle in this minimum at the end of inflation, a non-zero contribution to the cosmological constant is generically present. To ease notation, we define
\begin{eqnarray}
D&\equiv& N_{\mbox{\tiny axion}}\,,\\
n&\equiv& N_{\mbox{\tiny source}}\,,
\end{eqnarray}
and consider $n\gg D$ to guarantee a non-trivial landscape. If not stated otherwise, we use $n=125$ and vary $D$. 

\begin{figure}[htb]
    \centering
        \subfloat[ \label{fig:Vmin813:a}] {\includegraphics[width=0.48\textwidth, angle=0]{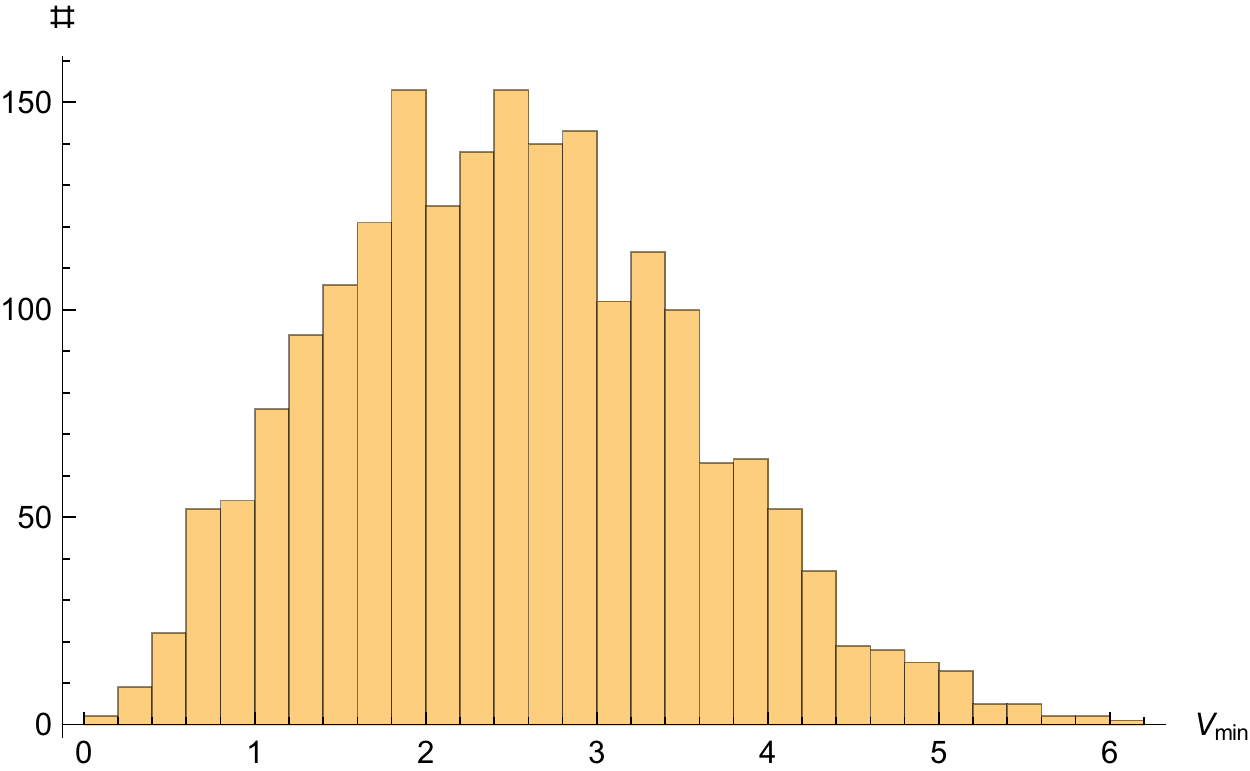}}
        \subfloat[ \label{fig:Vmin813:b}] {\includegraphics[width=0.48\textwidth, angle=0]{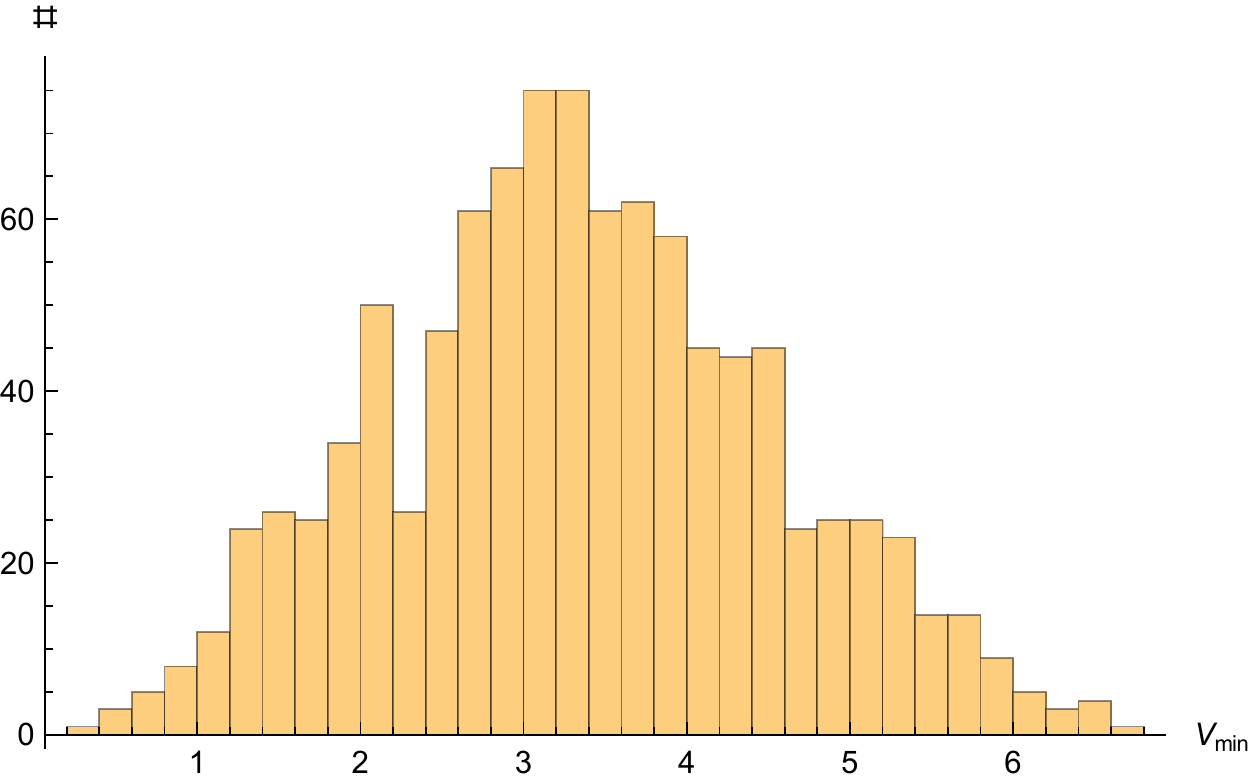}}\\
          \subfloat[ \label{fig:Vmin813:c}] {\includegraphics[width=0.48\textwidth, angle=0]{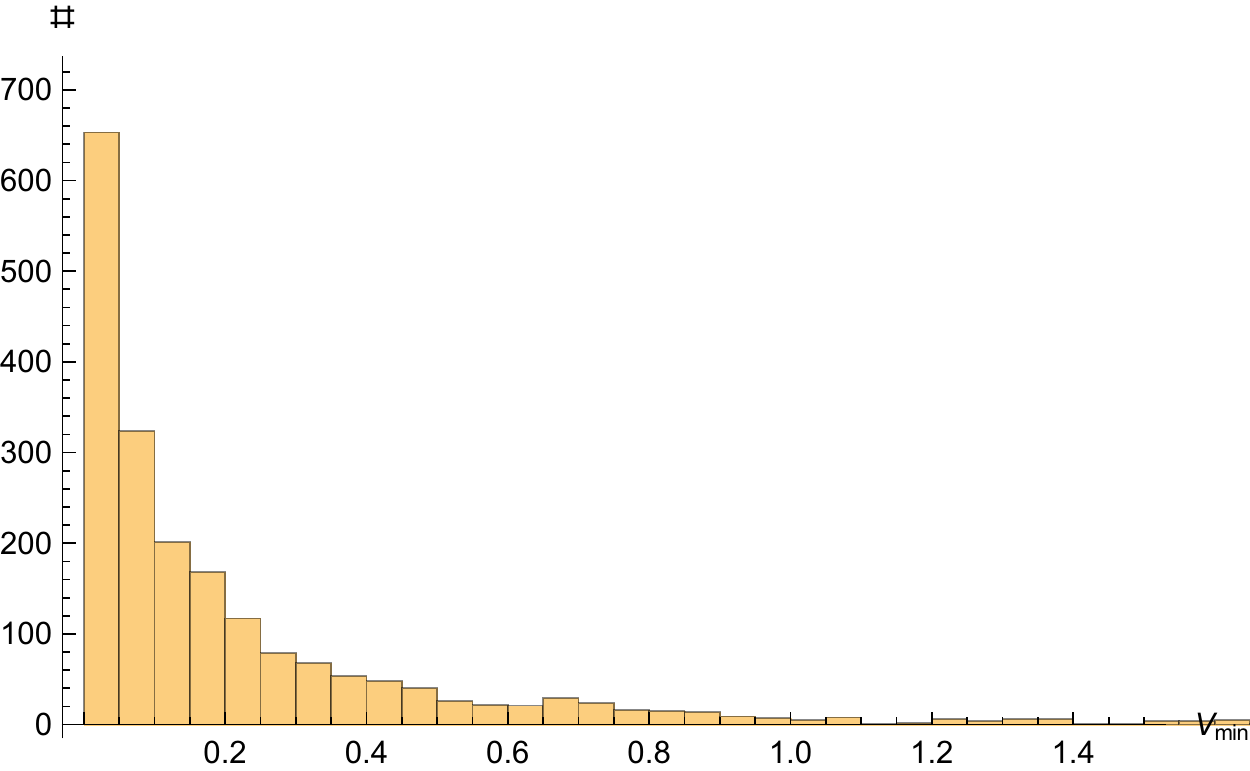}}
        \subfloat[ \label{fig:Vmin813:d}] {\includegraphics[width=0.48\textwidth, angle=0]{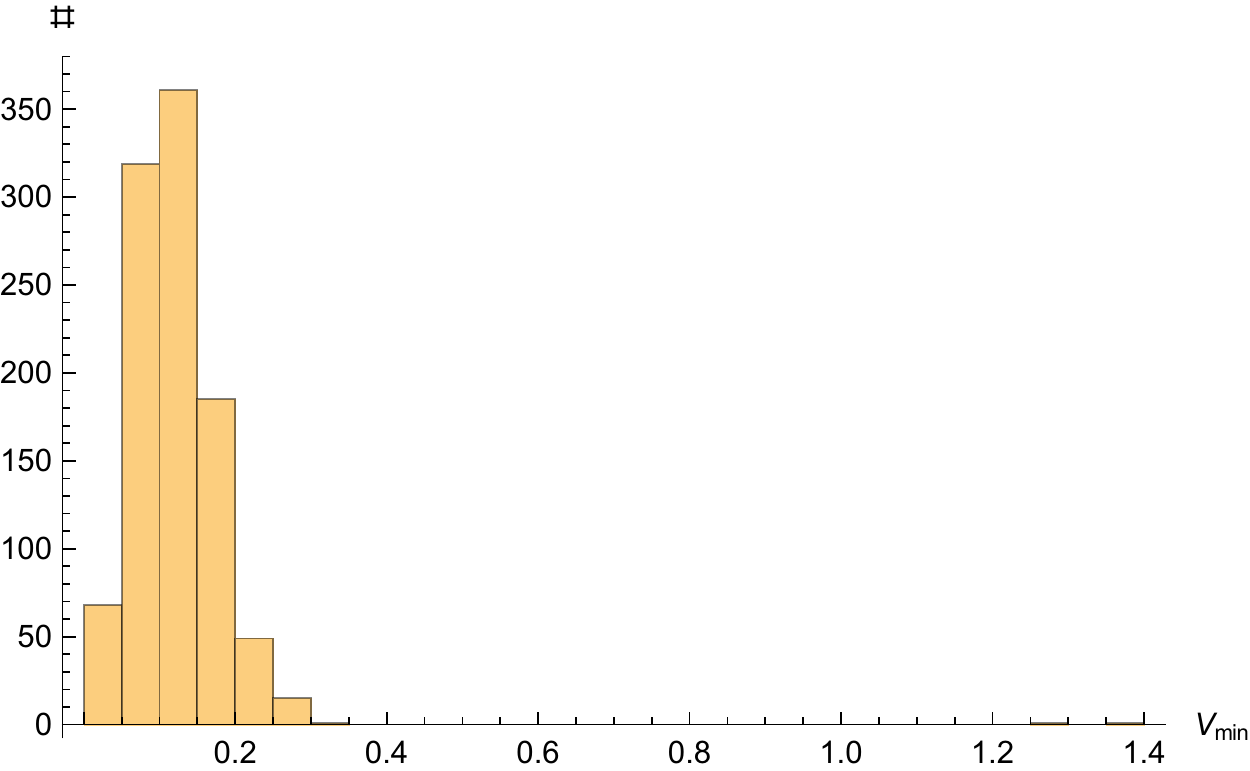}}
    \caption{Comparison to \cite{Higaki:2014mwa} for $n=13$ and $D=8$; all plots use the normalization $\sum \Lambda_i=n$, as in \cite{Higaki:2014mwa} (our normalization in the remainder of this article is $\sum \Lambda_i=1$): {a), $\Lambda_i=1$, counting all vacua with the same weight. The third insert of Fig.~1 in \cite{Higaki:2014mwa} is recovered. (b), $\Lambda_i=1$, minimas are found dynamically, see Sec.~\ref{sec:numerical studies} for details on the algorithm.  More minimas are found at larger values of $V$ in comparison to panel (a).  (c), $\Lambda_i$ are uniformly distributed, but all vacua are counted with equal weight. The distribution shifts to lower values of $V$. (d), Our approach up to the normalization of the potential: $\Lambda_i$ are uniformly distributed, minimas are found dynamically. Compared to panel (c), more minima are found at higher values of $V$. Histograms for case (d) show a large variability among different realisations of the potential due to the low value of $n$ in conjunction with a distribution for the $\Lambda_i$.}}
  \label{fig:Vmin813}
\end{figure}

Without loss of generality, we reabsorb the decay constants $f_j$ into the real valued mixing matrix $n_{ij}$, that is we formally set $f_j=1$;  in \cite{Higaki:2014mwa}, the mixing matrix was integer valued and the $f_j$ were kept explicit. If not stated otherwise, we draw the elements of $n_{ij}$ from a normal distribution with zero mean and standard deviation \begin{eqnarray}
\sigma_n=2\,,
\end{eqnarray}
truncated to the interval $[-3,3]$ and properly rescaled \footnote{We choose the same interval as in \cite{Higaki:2014mwa} for ease of comparison. There is not good a-priori theoretical reason to prefer this interval over another one.}.  The overall scale of the potential is set by the $\Lambda_i$, which we draw from a flat distribution over the interval [0,1].  We further renormalize the potential such that
\begin{eqnarray}
\sum_{i=1}^{n}\Lambda_i^4=1\,. \label{renormalization}
\end{eqnarray} 
For the relative phases, we choose a flat distribution over the interval $[0,2\pi)$. In \cite{Higaki:2014mwa}, $\Lambda_i\equiv \Lambda=1$ and $f_i=f$ were chosen, making their potentials less general, albeit still comparable to ours. To our knowledge, there is no theoretical reason to demand $\Lambda_i=\Lambda_j$ for $i\neq j$, except to keep the potential simple. Since the randomness of the potential is reduced by setting $\Lambda_i=\Lambda_j$ and histograms are affected by this simplification, see Fig.~\ref{fig:Vmin813}, we keep the $\Lambda_i$ as random variables. Additional differences of our approach in comparison to the one in \cite{Higaki:2014mwa} are the re-normalization of the potential according to (\ref{renormalization}) and the use of a dynamical algorithm to search for minima, see Sec.~\ref{sec:numerical studies}. A comparison for the case $n=13$ and $D=8$ is provided in Fig.~\ref{fig:Vmin813}:  using a distribution for the $\Lambda_i$ shifts the histograms to lover values, whereas the dynamical search for minima makes higher lying ones more likely to be found. We focus on the quantification of the latter effect in this publication. It should be noted that histogram (d) in Fig.~\ref{fig:Vmin813} shows large variability: some histograms show a peak while others don't. Furthermore, a Gaussian does not describe such histograms well. Both effects are due to the low value of $n$ used here. As a consequence, we use much larger values of $n$ in the remainder of this study.

Potentials constructed according to the above prescription are usually to steep to allow for slow roll inflation. As argued in \cite{Higaki:2014mwa}, one can increase the likelihood of inflation if the mixing matrix is chosen such that it is conductive for the KNP mechanism to be operational. To this end, define the matrix
\begin{eqnarray}
M_{\alpha\beta}\equiv \sum_{i=1}^{n}n_{i\alpha}n_{i\beta}\,,\label{massmatrix}
\end{eqnarray}
where $\alpha,\beta = 1,\dots,D$. Let $R$ be the ratio of the smallest eigenvalue of $M$ to the second smallest. As shown in \cite{Higaki:2014mwa}, demanding $R\ll 1$ increases the likelihood for a flat direction to be present in the landscape due to the KNP mechanism. We demand
\begin{eqnarray}
R<0.01 \label{conditionR}
\end{eqnarray}
 in our numerical studies if not stated otherwise.

\begin{figure}[htb]
    \centering
        \subfloat[$3D$ \label{fig:Vminhist:a}] {\includegraphics[width=0.3\textwidth, angle=-90]{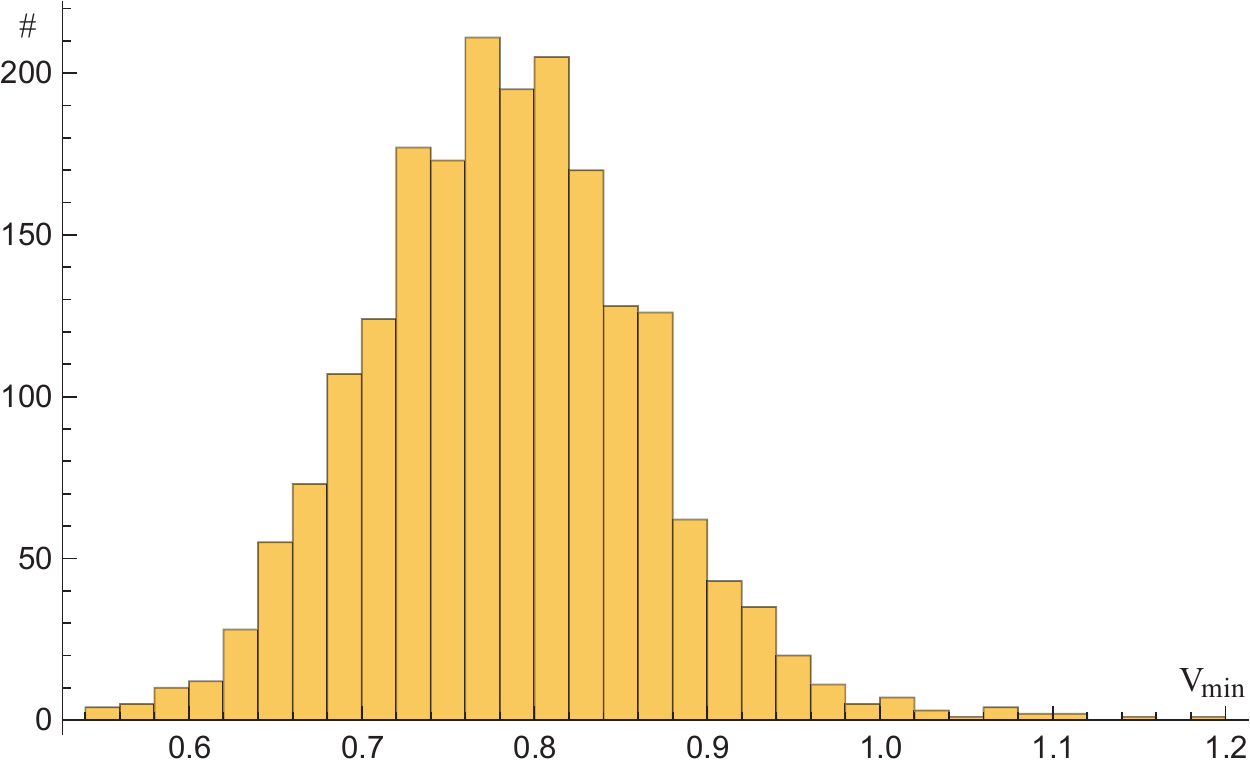}}
        \subfloat[$4D$ \label{fig:Vminhist:b}] {\includegraphics[width=0.3\textwidth, angle=-90]{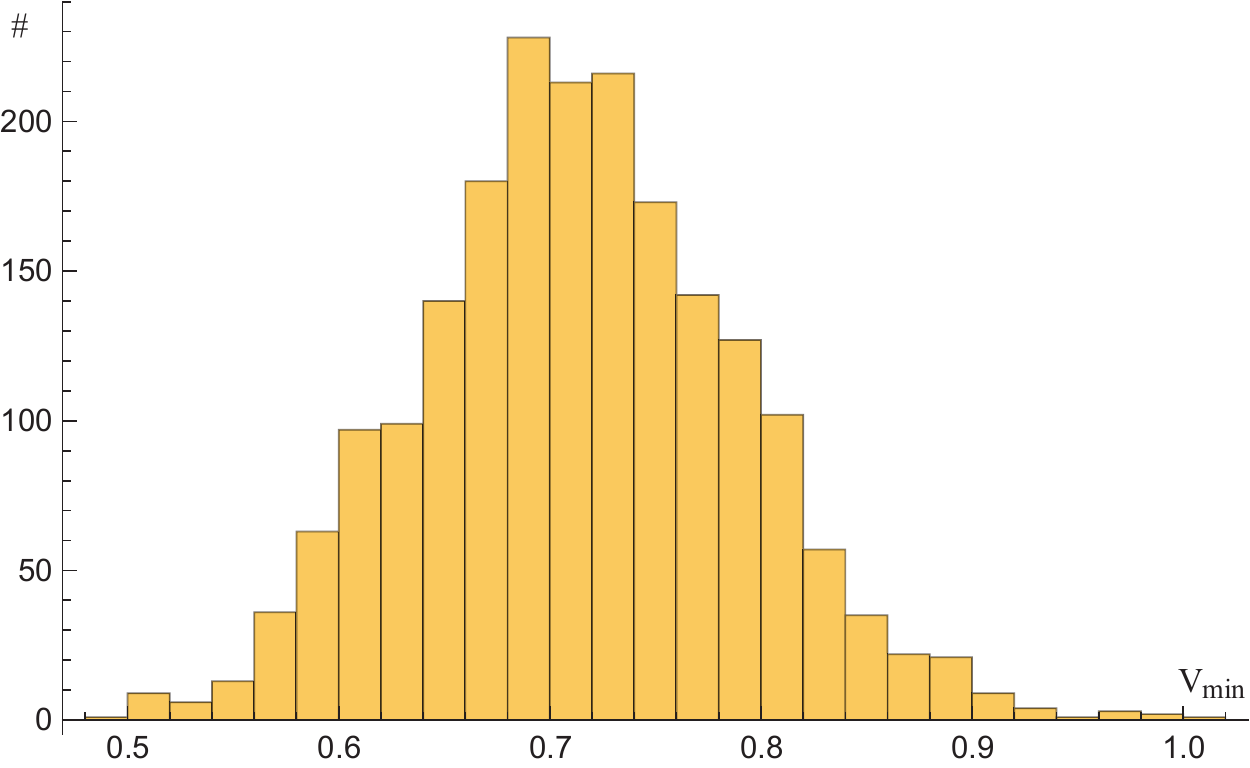}}\\
          \subfloat[$5D$ \label{fig:Vminhist:c}] {\includegraphics[width=0.3\textwidth, angle=-90]{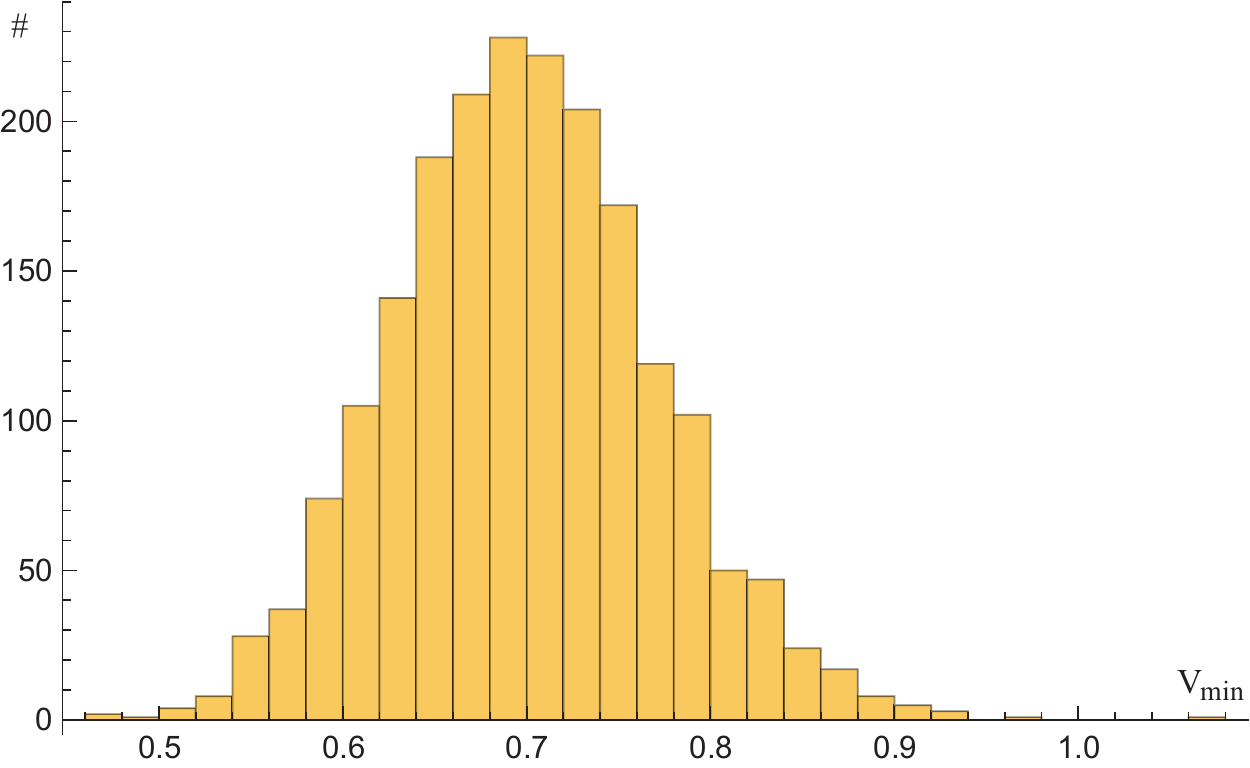}}
        \subfloat[$6D$ \label{fig:Vminhist:d}] {\includegraphics[width=0.3\textwidth, angle=-90]{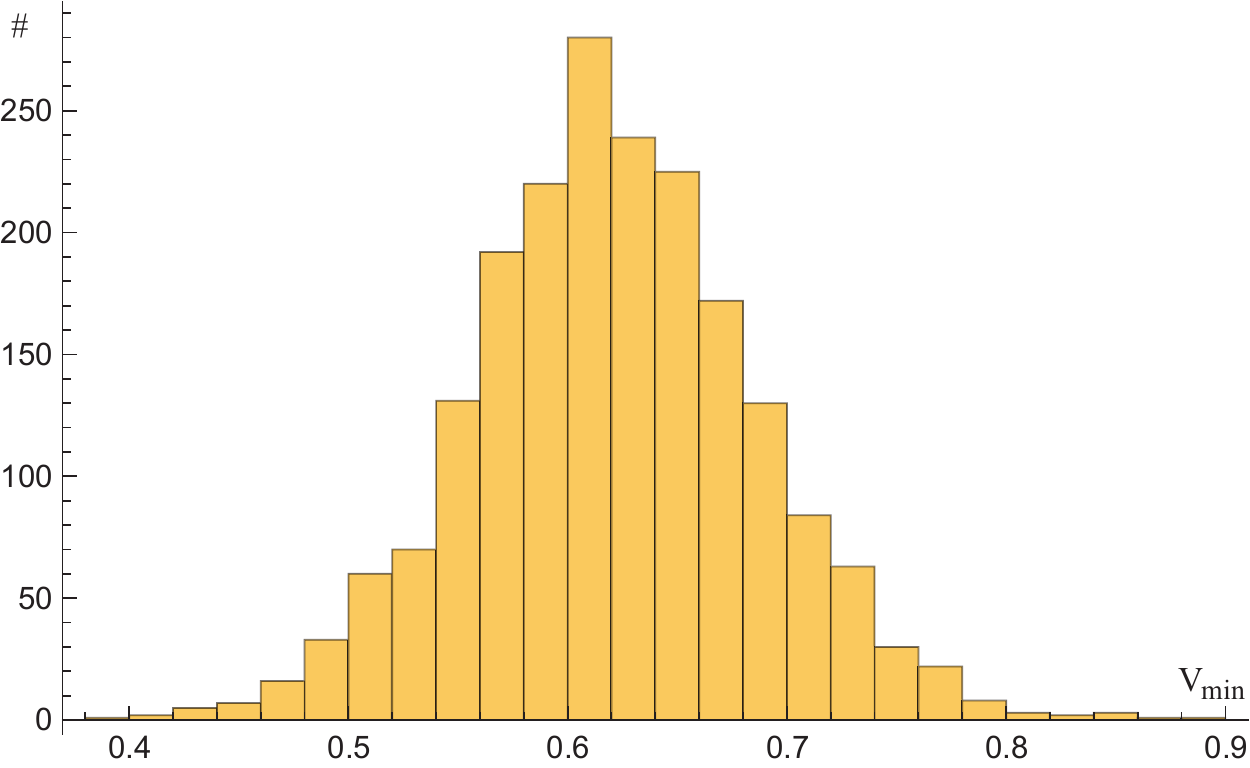}}
    \caption{{Exemplary distribution of the height at the dynamically reached minima $V_{\text{min}}$ for $D=3,4,5$ and $6$. For each $D$, minima are found according to the conditions (\ref{cond1}) and (\ref{cond2}) after solving the field equations (\ref{KGE}) in conjunction with (\ref{Friedmann}) within a single realization of the potential defined in (\ref{defpotential}); parameters are chosen according to Sec.~\ref{sec:modelparameters}. We identify $5000$ minima in each distribution. The peak of the histograms shifts to lower values as $D$ increases.}}
  \label{fig:Vminhist}
\end{figure}

\subsection{Numerical Computation: the Distribution of $ V_{\text{min}} $ for varying Dimension \label{sec:numerical studies}}

We construct an ensemble of axionic potentials according to (\ref{defpotential}) and model parameters in Sec.~\ref{sec:modelparameters}, including the condition on $R$ in (\ref{conditionR}). We would like to compute the height of those minima that are reached dynamically and investigate how the peak of this distribution changes as the dimensionality of field space is increased. 

To find minima, we first construct a potential. In this potential, we choose $5000$ random initial field values according to a flat distribution over the interval $[-10^{5},10^5]$ and let the axions evolve from rest, $\dot{\phi}_i=0$, by solving their equations of motion
\begin{eqnarray}
\ddot{\phi}_i+3H\dot{\phi}_i=-\frac{\partial V}{\partial \phi_i}\,,\label{KGE}
\end{eqnarray}
in conjunction with the Friedmann equation in a flat Universe
\begin{eqnarray}
3H^2=V+\sum_{i=1}^{D}\frac{\dot{\phi}^2_i}{2}\,,\label{Friedmann}
\end{eqnarray}
 where $H=\dot{a}/a$ is the Hubble parameter and a dot denotes a derivative with respect to cosmic time. We count all trajectories, regardless of whether they include an inflationary phase or not. To decide numerically if a minimum is reached, we demand 
 \begin{eqnarray} 
\dot{\phi_i}&<&0.01\,,\label{cond1}\\
\frac{1}{V}\frac{\partial V}{\partial \phi_i}&<&0.1\,\label{cond2}
 \end{eqnarray}
for $i=1,\dots,D$. While shallow regions and saddle point can be mistaken for minima, we checked that the resulting distributions of are not dominated by such misidentification. 

We expect the peak of the resulting distribution to shift closer to zero as $D$ is increased, as long as $D\ll n$, in line with the study of truncated Fourier series potentials in \cite{Battefeld:2012qx}. This is indeed qualitatively the case, see Fig.\ref{fig:Vminhist}. The resulting histograms are robust with respect to changes in the distribution of the $\Lambda_i^4$ (we checked a flat and a truncated Gaussian distribution) and the omission of the phases $\theta_i$. This independence of model parameters is expected due to the central limit theorem, as long as the number of independent random variables in the potential is large.

\begin{figure}[tb]
    \centering
        \subfloat[$\bar{V}_{\text{min}}$ \label{fig:VminoverD:a}] {\includegraphics[width=0.47\textwidth]{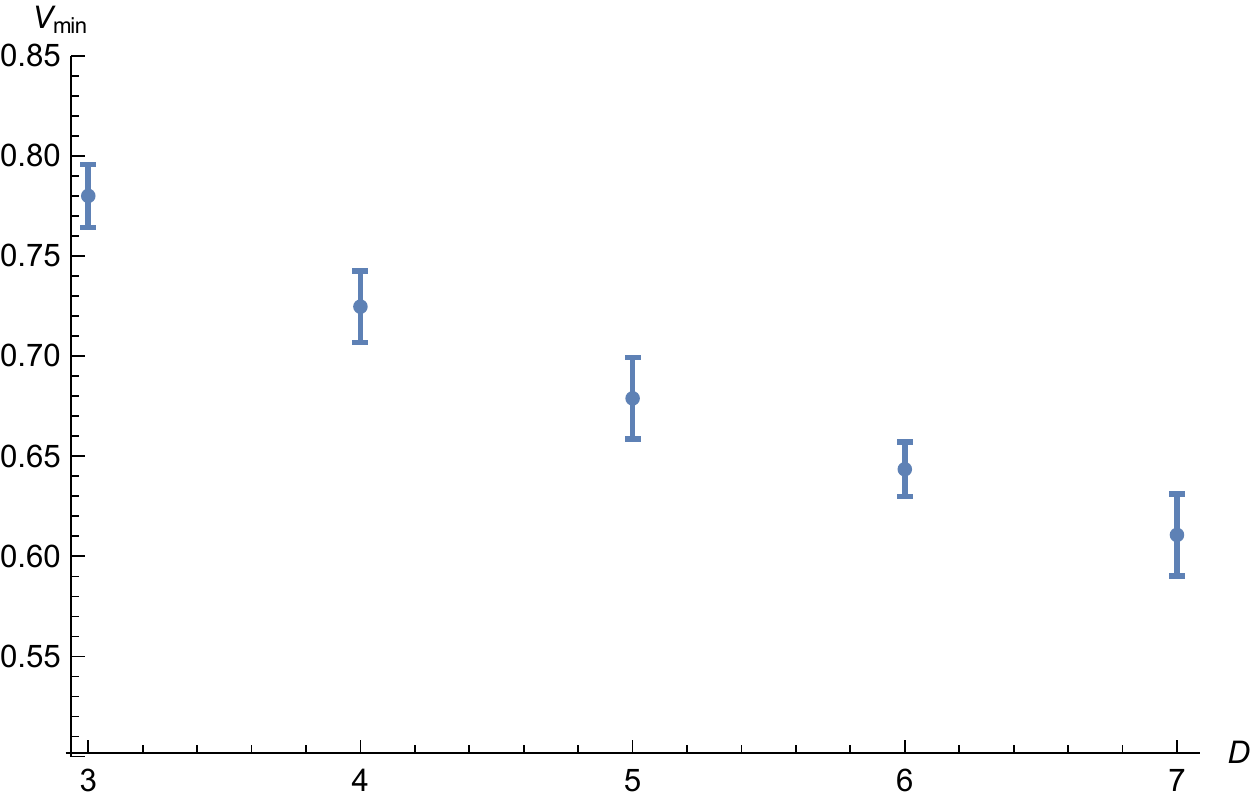}}
        \qquad %
        \subfloat[$\ln{\bar{V}_{\text{min}}}$ \label{fig:VminoverD:b}] {\includegraphics[width=0.47\textwidth]{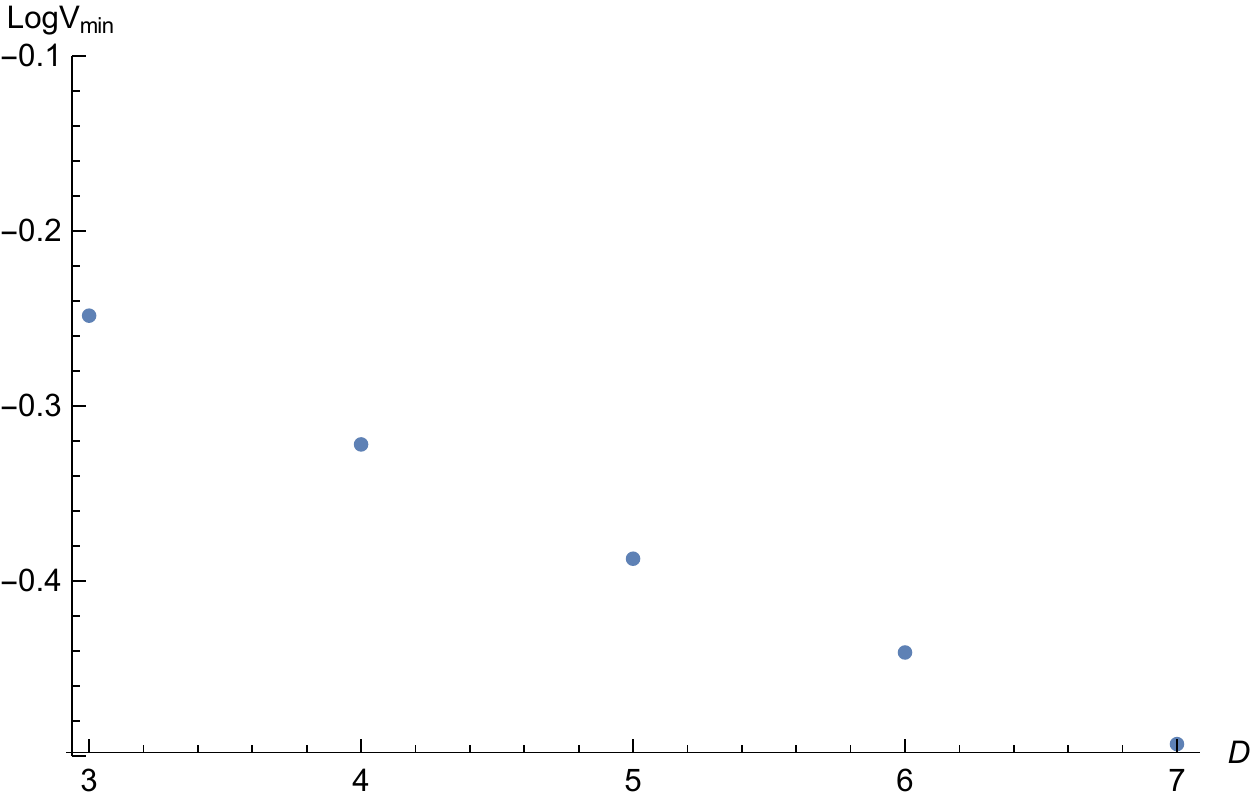}}
    \caption{{The peak in the histrogram of $500$ minima is averaged over $30$ realisations of the potential defined in (\ref{defpotential}) for $D=3,\dots, 7$ leading to the mean values $\bar{V}_{\text{min}}$ and $\ln{\bar{V}_{\text{min}}}$. Error bars in panel (a) indicate the $1\sigma$ variance. The resulting plot shows a clear trend, which may be fitted by a linear function with slope $-0.07$, see (\ref{fitlogV}).}}
  \label{fig:VminoverD}
\end{figure}

Keeping $D$ fixed, one may wonder how strongly the histograms vary from one realization to the next. To this end, we construct $30$ potentials for $D=3,\dots,7$ and identify $500$ dynamically reached minima for each potential. Noting the peak value for each histogram, we
can compute the mean $\bar{V}_{\text{min}}$ and variance, see Fig.~\ref{fig:VminoverD} and Tab.~\ref{tab:minima}: while some statistical scatter is clearly visible, the resulting variability is smaller than the observed trend.   The logarithm of $\bar{V}_{\text{min}}$ can be fitted well by a linear function
\begin{eqnarray}
\ln(\bar{V}_{\text{min}})&=&\text{const}+c D\,,\label{fitlogV}\\
\ln(\bar{V}_{\text{min}})&\approx& -0.074 - 0.061 D\,,
\end{eqnarray}
 for the narrow range of $D$ that are numerically accessible, but due to the limited reach in $D$, many other functions could be used as fitting functions as well. 

This trend in (\ref{fitlogV}) is in line with the theoretical expectation based on \cite{Battefeld:2012qx}. However, the analytic results in \cite{Battefeld:2012qx} can not be applied directly: the potential in (\ref{defpotential}) is lifted compared to the one in \cite{Battefeld:2012qx}, so that $V=0$ is the absolute minimum, and subsequently rescaled, such that $0\leq V \leq 2$. Furthermore, the coefficients are drawn from different distributions. The former could be accommodated by a rescaling of the results, but the latter may still have some impact, particularly for low $D$. Since the analytic result in \cite{Battefeld:2012qx} required a series of harsh approximations that were only tested for low $D$,  we postpone a detailed discussion of the above results and a comparison with \cite{Battefeld:2012qx} to Sec.~\ref{sec:inflation}.

\begin{table}
\begin{center}
    \begin{tabular}{|c|c|c|c|c|c|} \hline  
    \rule{0pt}{12pt} $D$ & 3 & 4 & 5 & 6 & 7 \\ [0ex] \hline \hline
    \rule{0pt}{18pt} $\bar{V}_{\text{min}}$ & 0.7795 & 0.7247 & 0.6789 & 0.6435  & 0.6107  \\ [1ex] \hline 
    \rule{0pt}{18pt} $\ln{\bar{V}_{\text{min}}}$ & -0.249 & -0.322 & -0.387 & -0.441 & -0.493  \\ [1ex] \hline    \end{tabular}
\caption{The peak in the histrogram of $500$ minima is averaged over $30$ realisations of the potential defined in (\ref{defpotential}) for $D=3,\dots, 7$ leading to the mean values $\bar{V}_{\text{min}}$ and $\ln{\bar{V}_{\text{min}}}$. See Fig.~\ref{fig:VminoverD} for the accompanying variance.   \label{tab:minima}}
\end{center}
\end{table}

\section{Properties of Random Axionic Potentials\label{sec:Properties}}
Given the random axionic potential in (\ref{defpotential}),
we would like to explain analytically the shape of the histograms in Fig.~\ref{fig:Vminhist} as well as the the dependence of $\bar{V}_{\text{min}}(D)$ in (\ref{fitlogV}). To this end, we aim to estimate first the number of critical points $n_c$ that are encountered on average as fields traverse a distance $\delta V$ down the potential. In combination with the statistical properties of the Hessian, one should be able to compute the ensemble average of the peaks' position in the histograms in Fig.~\ref{fig:Vminhist}, similar to the approach taken in \cite{Battefeld:2012qx}. Furthermore, in the large $D$ limit, one should be able to employ random matrix theory to attain analytic results.

To make the problem more tractable analytically, we make the potential less random: in (\ref{defpotential}) the mixing parameters $n_{ij}$ were random. Here, we set them deterministically to 
\begin{eqnarray}
n_{ij}\equiv \dfrac{3J_j}{\tilde{n}}\,,
\end{eqnarray}
 where $J_j$ runs from 1 to $\tilde{n}$ and
\begin{eqnarray}
\tilde{n}\equiv n^{1/D}\,.
\end{eqnarray} 
Evidently, this definition makes sense only if $\tilde{n}$ is an integer, that is only for certain combinations of $n$ and $D$, say $n=125$ and $D=5$ so that $\tilde{n}=3$. As we shall see, the final results are independent of $\tilde{n}$ so that this restriction can be lifted in retrospect\footnote{We replace sums by integrals at some point, so that non-integer $\tilde{n}$ do not pose a problem any more.}.
By construction, elements in the mixing matrix cover the interval $(0,\dots,3]$ evenly instead of randomly and they are independent of the field index $i=1,\dots, D$. We checked numerically that restricting the range to $(0,\dots,3]$ from $[-3,\dots,3]$ has no effect on the resulting histograms, in line with the robustness of the results with respect to the omission of the phase factors $\theta_i$. However, it should be noted that the above simplification renders all entries of the mass matrix in (\ref{massmatrix}) identical so that the KNP mechanism is not operational. 

Next, we alter the notation for the $n=\tilde{n}^D$ independent random variables $\Lambda_i^4$  to $\Lambda_{J_1,...,J_D}$. To enable the analytic computation of subsequent integrals, we further switch the flat distribution of the $\Lambda_i$ to a normal distribution for the $\Lambda_{J_1,\dots,J_D}$ with mean
\begin{eqnarray}
\mu_{\Lambda}\equiv \frac{1}{\tilde{n}^D}=\frac{1}{n}
\end{eqnarray}
and standard deviation
 \begin{eqnarray}
 \sigma_{\Lambda}&\equiv& \frac{a}{\tilde{n}^D}\,,\\
 a&\equiv& 0.1\,.
\end{eqnarray} 
We chose $\mu_{\Lambda}\equiv 1/n$, so that our potential remains nearly normalized for different $n$ and $D$, that is 
\begin{eqnarray}
\sum_{J_1,\dots, J_D=1}^{\tilde{n}} \Lambda_{J_1,\dots,J_D}\approx 1\,. \label{normalization}
\end{eqnarray}
 $a$ should be a small number so that negative values of  $\Lambda_{J_1,\dots,J_D}$ are rare. We checked that different choices for $a$ don't change our results qualitatively as long as $a\ll 1$.
 
Given these simplifications and definitions, the potential in (\ref{defpotential}) reads
\begin{equation}
 V=\sum_{J_1,\dots,J_D=1}^{\tilde{n}} \Lambda_{J_1,\dots,J_D}\bigg(1-\cos\big(\sum_{j=1}^D \frac{3J_j}{n} \phi_j+\theta_{J_1,\dots,J_D}\big)\bigg)\,.\label{simplifiedpotential}
\end{equation}
This potential is quite similar to the one investigated in \cite{Battefeld:2012qx}, so that similar techniques to attain analytic results can be employed as well. While our simplifications reduce the randomness and alter the distribution of random variables, we expect our subsequent results pertaining to the distribution of dynamically reached minima to remain valid as long as the number of random parameters entering the potential remains large, that is in the limit $n\gg D\gg 1$, essentially due to the central limit theorem.\footnote{We checked numerically that the histograms and thus $\bar{V}_{\text{min}}(D)$ obtained from this simplified potential are consistent with the numerical results of Sec.~\ref{sec:numerical studies} for the values of $n,D$ used in this article. It should be noted that the case studies in \cite{Higaki:2014mwa} have such low values of $n$ and $D$ that our subsequent analytical results do not apply there.}

To reiterate, our goal is to estimate the number of critical points as fields roll down the potential, see Sec.~\ref{sec:computenc}, and compute the likelihood that a critical point is a minimum, see Sec.~\ref{subsec:whyrmtfails}, so that we can compute the likely height of the final resting place.

\subsection{The PDF of V}{\label{Sec:pdfpotential}}
The potential in (\ref{simplifiedpotential}) is the sum of products of random variables, the $\Lambda_{J_1,\dots,J_D}$ and 
\begin{eqnarray}
\tilde{x}_{J_1,\dots,J_D}\equiv \bigg(1-\cos\big(\sum_{j=1}^D \frac{3J_j}{n} \phi_j+\theta_{J_1,\dots,J_D}\big)\bigg)\,.\label{deftildex}
\end{eqnarray}
Since the distribution of the latter is much broader than the former and reasonably flat, we may approximate the PDF of the product by the distribution of the $\Lambda_{J_1,\dots,J_D}$. As a consequence, the potential at a random point is to a good approximation a normally distributed random variable with mean and variance given by
\begin{eqnarray}
\mu_V&=&1\,,\\
\sigma_V&=&\frac{a}{\sqrt{n}}\,. \label{varianceV}
\end{eqnarray}
The numerical pre-factor in the variance is caused by the redefinition of the potential in (\ref{simplifiedpotential}) as well as replacing the $\tilde{x}_{J_1,\dots,J_D}$ by their expectation values, but the scaling $\sigma_{V}\propto 1/\sqrt{n}$ is robust and can be understood as follows: without renormalizing the potential, each term in the sum adds a random step, with even chance to be above or below one. The expectation value of the square-distance to the mean of this random walk scales with $\sqrt{n}$. Since we rescaled the potential by a factor of $1/n$, the square-distance scales as $1/\sqrt{n}$, which in turn is proportional to $\sigma_{V}$. This feature has  consequences for the feasibility of inflation on such potentials, see Sec.~\ref{inflation}.

\subsection{Estimation of the Number of Critical Points $n_c$ \label{sec:computenc}}
In this section, we estimate the number of critical points $n_c$ that are encountered if a potential difference of $\delta V$ is traversed.  
\subsubsection{PDF of the Potential's Gradient\label{sec:PDFpotentialgradient}}
To estimate $n_c$, we need the probability density function (PDF) of the partial derivative
\begin{equation}
V_k\equiv \frac {\partial V}{\partial \phi_k}=\sum_{J_1,\dots,J_D=1}^{\tilde{n}} \Lambda_{J_1,\dots,J_D}\frac{3J_k}{\tilde{n}} \sin \bigg( \sum_{j=1}^D \frac{3J_j}{\tilde{n}} \phi_j+\theta_{J_1,\dots,J_D}\bigg)
\end{equation}
to compute the PDF of the gradient's absolute magnitude,
\begin{eqnarray}
V^\prime\equiv \sqrt{\sum_{k=1}^{D}V_k^2}\,.\label{defvprime}
\end{eqnarray}
To ease notation, we define
\begin{eqnarray}
x_{J_1,\dots,J_D}\equiv 1+\sin \bigg( \sum_{j=1}^D \frac{3J_j}{\tilde{n}} \phi_j+\theta_{J_1,\dots,J_D}\bigg)\,,\label{defx}
\end{eqnarray}
so that the partial derivative of the potential reads
\begin{equation}
 V_k=\sum_{J_1,\dots,J_D=1}^{\tilde{n}} \Lambda_{J_1,\dots,J_D}\frac{3J_k}{\tilde{n}} (x_{J_1,\dots,J_D}-1)\,. 
 \end{equation}
Since the argument of the $\sin$-function at a given point is a random variable, $x_{J_1,\dots,J_D}$ is a random variable as well, which is symmetric around its mean $1$. Evidently, $V_k$ is a random variable with zero mean. We approximate the distribution of  $x_{J_1,\dots,J_D}$ by a flat one over the interval $[0,2]$. Furthermore, defining
\begin{equation}
 b_{J_1,\dots,J_D}\equiv\Lambda_{J_1,\dots,J_D}-\mu_{\Lambda}\in \mathcal{N}(0,\sigma_{\Lambda})\,,
 \end{equation}
the derivative of the potential becomes
\begin{eqnarray}
 V_k&=&\sum_{J_1,\dots,J_D=1}^{\tilde{n}} \frac{3J_k}{\tilde{n}} (b_{J_1,\dots,J_D}+\mu_{\Lambda})\cdot(x_{J_1,\dots,J_D}-1)\\
 \nonumber  &=&\frac{3}{\tilde{n}} \sum_{N_p} \Big(\sum^{\tilde{n}}_{J_{k}=1} b_{J_1,\dots,J_D} \cdot x_{J_1,\dots,J_D}\cdot  J_k -\sum^{\tilde{n}}_{J_{k}=1} b_{J_1,\dots,J_D}\cdot J_k +\sum^{\tilde{n}}_{J_{k}=1} \mu_{\Lambda} \cdot x_{J_1,\dots,J_D}J_k-\sum^{\tilde{n}}_{J_{k}=1} \mu_{\Lambda}\cdot  J_k \Big) \\\label{Vkfourterms}
\end{eqnarray}
Here, $N_p$ stands for the number of permutations of the $J_j (j\neq k)$ under which $J_k$ doesn't change, i.e.
\begin{equation}
 N_p=\tilde{n}^{D-1}\,,
 \end{equation} 
 and we  introduced the shorthand notation
\begin{equation}
\sum_{N_p}(\dots)\equiv \sum_{J_1,\dots,J_{k-1},J_{k+1},\dots,J_D=1}^{\tilde{n}}(\dots)\,.\label{shorthand}
\end{equation}
We proceed by calculating the variance for each of the four terms in (\ref{Vkfourterms}) separately, keeping in mind that the mean of $V_k$ is zero.

\paragraph{The 1'st and 2'nd Term:}
The PDF of the product
\begin{equation}
h=b_{J_1,\dots,J_D}\cdot x_{J_1,\dots,J_D}\label{defh}
\end{equation}
where $x_{J_1,\dots,J_D}$ has a flat distribution and $b_{J_1,\dots,J_D}$ a Gaussian one, is given by \begin{equation}f(h)=\dfrac{1}{c} \operatorname{Ei}\left(-\frac{h^2}{8\pi \sigma_{\Lambda}}\right)\,,
\end{equation} 
where $c$ is a normalization constant and $\operatorname{Ei}$ the exponential Integral. We approximate this PDF by a normal distribution with zero mean and the same variance as $b_{J_1,\dots,J_D}$, i.e. $\sigma_\Lambda$. This approximation is equivalent to replacing $x_{J_1,\dots,J_D}$ by its mean value of $1$ in (\ref{defh}), that is, we ignore the $\sin$-function.  In this approximation, the variance of the first term in (\ref{Vkfourterms}) becomes
\begin{eqnarray}
\sigma_1^2&=&\frac{9}{\tilde{n}^2}N_p  \sigma_{\Lambda}^2  \sum^{\tilde{n}}_{J_k=1} J_k^2  \\
&=&\frac{9}{\tilde{n}^2}\tilde{n}^{D-1}\frac{a^2}{\tilde{n}^{2D}} \left(\frac{(\tilde{n}+1)^3}{3}+\frac{(\tilde{n}+1)^2}{2}+\frac{\tilde{n}}{6}+\frac{1}{6}\right)\\
&\approx & \frac{3a^2}{\tilde{n}^D}=\frac{3a^2}{n}\,,
\end{eqnarray}
where we kept the leading order term in $\tilde{n}$ only. Similarly, the variance of the second term becomes
\begin{eqnarray}
\sigma_2^2 =n \sigma_{\Lambda}^2=\frac{a^2}{n}
\end{eqnarray}
since $N_p\tilde{n}=\tilde{n}^D=n$\,.

\paragraph{The 3'rd and 4'th Term:}
Consider
\begin{eqnarray}
b\equiv \sum^{\tilde{n}}_{J_{k}=1} x_{J_1,\dots,J_D} \cdot  J_k\ =\sum_{i=1}^{\tilde{n}} 2iy_i .
\end{eqnarray}
where $y_i$ are independent and identically uniformly distributed, $U(0,1)$, random variables. According to  App.~\ref{sec:distribution}, equation (\ref{a2}), the variance of this sum is
\begin{equation}
\sigma_{b}^2=C \tilde{n}^2(\tilde{n}+1)\,,
\end{equation}
with $C\approx 0.11$. As a consequence, the variance of the third term in (\ref{Vkfourterms}) is
\begin{eqnarray}
 \sigma_{3}^2&=&\mu_{\Lambda}^2  N_p  \frac{9}{\tilde{n}^2} \cdot \sigma_b^2
 =9C\frac{1}{\tilde{n}^D}\frac{\tilde{n}^2+\tilde{n}}{\tilde{n^2}} 
\approx \frac{0.99}{\tilde{n}^D}\approx \frac{1}{n}\,,
 \label{sigma3gradient}
\end{eqnarray}
where we kept the leading order term in the last step. The variance of the forth term is zero, since it does not contain any random variable.

Since $V_k$ is the sum of three approximately Gaussian random variables and a constant, its variance becomes
\begin{eqnarray}
\sigma_{V_{k}}^2= \sum_{i=1}^3\sigma_i^2 \approx \frac{9C+4a^2}{n}\approx \frac{1}{n}\label{sigmaVk}
\end{eqnarray}
to leading order in $\tilde{n}$; in the last step, we neglected $a^2=0.01 \ll 1$ and used $9C\approx 1$.
Note that the constant proportionalality factor of order one in (\ref{sigmaVk}) can be reabsorbed by the free parameter $\beta$ below, so that it is not crucial for us to keep any sub-leading terms. Hence, we approximate the PDF of $V_k$ by
\begin{equation}
f(V_k)=\frac{1}{\sqrt{2\pi\sigma_{V_{k}}^2}}\exp\left(-\frac{V_k^2}{2\sigma_{V_{k}}}\right)\,.
\end{equation}
Consequently, the absolute magnitude of the potential's gradient in (\ref{defvprime})
obeys a $\chi$-distribution,
\begin{eqnarray}
f(V^\prime)
=\frac{V'^{D-1}}{\sigma^D 2^{\frac{D-2}{2}} \Gamma \left( \frac{D}{2} \right)}\exp \left(-\frac{V'^2}{2 \sigma^2}\right)\,,\label{pdfofvprime}
\end{eqnarray}
with
\begin{eqnarray}
\sigma =\sigma_{V_k}\approx \frac{1}{\sqrt{n}}\,.
\end{eqnarray}
In the large $n,D$-limit, the expectation value of $V^{\prime}$ can be approximated by
\begin{eqnarray}
\overline{V^{\prime}}=\int_{0}^{\infty}V^\prime f(V^\prime)\,\text{d}V^\prime&=&\sigma\sqrt{2}\frac{\Gamma\left(\frac{D+1}{2}\right)}{\Gamma\left(\frac{D}{2}\right)}
\approx \sqrt{D/n}\,,\label{expectationvalueVprime}
\end{eqnarray} 
where we used the identity $\Gamma(x/2)\Gamma((x+1)/2)=\sqrt{\pi}\Gamma(x)/2^{x-1}$ as well as the large argument limit $\Gamma(x)\approx \sqrt{2\pi}x^{x-1/2}\exp(-x)$ for $x\gg 1$. 

\subsubsection{Potential Difference to the Nearest Critical Point and Estimate of $n_c$\label{subsec:difftocri}}
We would like to estimate the potential difference $\Delta V$ between a random initial point with slope $V^\prime$ defined in (\ref{defvprime}) and the nearest critical point with $V^\prime_{\text{critical}}=0$. Since our potential is relatively smooth, we expect this difference to be bigger, the larger the initial slope is. Thus, we make the Ansatz
\begin{eqnarray}
\Delta V(V^\prime)=\frac{V^\prime}{\beta \sqrt{D}}\,,\label{Ansatzbeta}
\end{eqnarray}
where $\beta$ is a constant and we took out the expected scaling with the dimensionality of field space. Given the PDF of $V^\prime$ in (\ref{pdfofvprime}), we can compute the expectation value of $\Delta V$ to
\begin{eqnarray}
\overline{\Delta V} &=&\int_{0}^{+\infty} \Delta V(V') f(V') \, \mathrm{d}V'\\
&=& \int_{0}^{+\infty} \frac{V^\prime}{\beta \sqrt{D}} f(V') \, \mathrm{d}V'\,.
\end{eqnarray}
Evaluating the Gaussian integral yields
\begin{eqnarray}
\overline{\Delta V}&=&\sqrt{\frac{2}{D}}\frac{\sigma}{\beta}\frac{\Gamma \left( \frac{D+1}{2} \right)}{\Gamma \left( \frac{D}{2} \right)}\,,\label{deltav}
\end{eqnarray}
where $\sigma^2=1/n$ according to (\ref{sigmaVk}). In the large $D$ limit, this potential difference becomes
\begin{eqnarray}
\overline{\Delta V}\approx \sqrt{\frac{1}{n\beta}}\,,\label{deltavlarged}
\end{eqnarray}
where we used the large argument limit of the Gamma-function.
 We tested the Ansatz in (\ref{Ansatzbeta}) numerically for varying $D$ and $\tilde{n}$ and found that $\beta$ is indeed approximately a constant of order one,
\begin{eqnarray}
 \beta \approx 1.9 \label{valueofbeta}\,,
 \end{eqnarray} 
 see Table \ref{table:beta}, which justifies our Ansatz. 
 
\begin{table} 
\begin{center}
\begin{tabular}{| l | l |} \hline
\rule{0pt}{15pt} $D=2\,, \tilde{n}=5$ & $\bar{\beta}$ =2.06 \\ \hline
\rule{0pt}{15pt} $D=2\,, \tilde{n}=7$ & $\bar{\beta}$ =1.99 \\ \hline
\rule{0pt}{15pt} $D=3\,, \tilde{n}=4$ & $\bar{\beta}$ =1.83 \\ \hline
\rule{0pt}{15pt} $D=3\,, \tilde{n}=5$ & $\bar{\beta}$ =2.09 \\ \hline
\rule{0pt}{15pt} $D=3\,, \tilde{n}=6$ & $\bar{\beta}$ =1.87 \\ \hline
\rule{0pt}{15pt} $D=4\,, \tilde{n}=5$ & $\bar{\beta}$ =1.86 \\ \hline
\rule{0pt}{15pt} $D=4\,, \tilde{n}=7$ & $\bar{\beta}$ =1.72 \\ \hline
\end{tabular}
\caption{To test the Ansatz in (\ref{Ansatzbeta}), we compute $\beta$ numerically, while varying the number of fields $D$ and the number of sources $n=\tilde{n}^D$. $\bar{\beta}$ is the average value of $\beta$ for $3$ numerical runs.}
\label{table:beta}
\end{center}
\end{table}  
 
Given $ \overline{\Delta V}$, we can estimate  the number of critical points $n_c$ that are encountered if traversing a potential difference of $\delta V$  by
\begin{eqnarray}
 n_c(\delta V)&=&\dfrac{\delta V}{ \overline{\Delta V}}=\beta\,\delta V\sqrt{\frac{D}{2n}} \frac{\Gamma \left( \frac{D}{2} \right)}{\Gamma \left( \frac{D+1}{2} \right)}\\
 &\approx&\delta V\sqrt{\beta n}\,.\label{ncapprox}
\end{eqnarray}

\subsection{The Hessian\label{subsec:Hessian}}
We are interested to compute the ratio of minima to critical points, particularly how this ratio changes with the dimensionality of field space. Thus, we need to compute the PDF of the Hessian. 

Consider a random critical point $\phi_c$ at a height $V(\phi_c) \equiv V_{\text{c}}$. In this section a subscript $c$ is not a free index, but denotes a quantity evaluated at a critical point. From the definition of the Hessian and the potential in (\ref{defpotential}) we get
\begin{eqnarray}
H_{kl}\Big|_{V_{\text{c}}}&=&\frac{\partial^2 V}{\partial \phi_k \phi_l}\Big|_{V_{\text{c}}} \\
&= & -\frac{9}{\tilde{n}^2}V_{\text{c}}+\tilde{H}_{kl},
\end{eqnarray}
with
\begin{eqnarray}
\tilde{H}_{kl} \equiv \frac{9}{\tilde{n}^2}\bigg( \sum_{J_1,\dots,J_D=1}^{\tilde{n}}\Lambda_{J_1,\dots,J_D}+\sum_{J_1,\dots,J_D=1}^{\tilde{n}}\Lambda_{J_1,\dots,J_D} \cos(\sum_{j=1}^D \frac{3J_j}{\tilde{n}} \phi_j+\theta_{J_1,\dots,J_D})(J_k J_l-1)\bigg)\,.
\end{eqnarray}
We extracted the term proportional to $V_{\text{c}}$ to highlight the deterministic dependence of the Hessian, and thus the likelihood of a minimum, on the height of the critical point $V_{\text{c}}$. This dependence is crucial to retain, as first pointed out in \cite{Battefeld:2012qx}.
The random contribution $\tilde{H}_{kl}$ has a mean of
\begin{eqnarray}
\mu_{\tilde{H}}=\frac{9}{\tilde{n}^2}\overline{\sum_{J_1,\dots,J_D=1}^{\tilde{n}} \Lambda_{J_1,\dots,J_D}}\approx \frac{9}{\tilde{n}^2} \,,
\end{eqnarray}
where we used the approximate normalization of the potential in (\ref{normalization}). Readers not interested in the derivation of the variance may jump to Sec.~\ref{sec:summaryhessian} for the result.

To compute the standard deviation of $\tilde{H}_{kl}$,
we rewrite $\tilde{H}_{kl}$ as
\begin{eqnarray}
\nonumber \tilde{H}_{kl}&= & \frac{9}{\tilde{n}^2} \bigg(-\sum_{J_1,\dots,J_D=1}^{\tilde{n}}\Lambda_{J_1,\dots,J_D}(J_kJ_l-2) +\sum_{J_1,\dots,J_D=1}^{\tilde{n}} b_{J_1,\dots,J_D} x_{J_1,\dots,J_D}(J_kJ_l-1) \\
&&+\sum_{J_1,\dots,J_D=1}^{\tilde{n}}\mu_{\Lambda}x_{J_1,\dots,J_D}{(J_kJ_l-1)} \bigg)\,,
\end{eqnarray}
where we defined the random variables\footnote{While the variable $\tilde{x}$ and $x$ in (\ref{deftildex}) and (\ref{defx}) have different definitions, they have the same probability distribution if viewed as random variables. To avoid an overly complicated notation, we denote the analogous variable below by $x$, even though it is again a different variable with the same probability distribution.}
\begin{eqnarray}
b_{J_1,\dots,J_D} &\equiv& \Lambda_{J_1,\dots,J_D}-\mu_{\Lambda}\,,\\
 x_{J_1,\dots,J_D} &\equiv& \cos\big(\sum_{j=1}^D \dfrac{3J_j}{\tilde{n}} \phi_j+\theta_{J_1,\dots,J_D}\big)+1\,.
 \end{eqnarray}
 The $b_{J_1,\dots,J_D}$ are normal distributed with zero mean and variance $\sigma_{\Lambda}$ and, similar to Sec.~\ref{sec:PDFpotentialgradient}, we approximate the PDF of $x_{J_1,\dots,J_D}$ by a uniform distribution on the interval $[0,2]$. To keep the notation clean, we suppress the multi-index $J_1,\dots, J_D$ on $b$ and $x$ in the following. Furthermore, it is useful to compute the standard deviation of the diagonal and off-diagonal elements of $\tilde{H}_{kl}$ separately.
 
\subsubsection{Diagonal Elements of $\tilde{H}_{kl}$}
For $ k = l $, we have
\begin{equation}
\tilde{H}_{kk}=\frac{9}{{\tilde{n}}^2} \sum_{N_p} \Big(-\sum^{\tilde{n}}_{J_k=1}\Lambda_{J_1,\dots,J_D}(J_k^2-2)+\sum^{\tilde{n}}_{J_k=1}bx(J_k^2-1) +\sum^{\tilde{n}}_{J_k=1}\mu_{\Lambda}x(J_k^2-1)\Big).
\label{htildedia}
\end{equation}
where we used the shorthand notation defined in (\ref{shorthand}).

The first term in (\ref{htildedia}) is a sum of Gaussians with identical variance and constant coefficients, so that it is also a Gaussian with variance
\begin{eqnarray}
\sigma_1&=&\frac{9}{{\tilde{n}}^2}\sqrt{N_p}\sigma_\Lambda\sqrt{\sum_{J_k=1}^{\tilde{n}}(J_k^2-2)^2}\\
&\approx& \frac{9a}{\sqrt{5n}}
\end{eqnarray}
where we used $\sigma_{\Lambda}=a/n$, $N_p=\tilde{n}^{D-1}$, $\tilde{n}^D=n$ and kept only the leading order term  in $\tilde{n}$ in the last step.

The second term in (\ref{htildedia}) contains the product of a Gaussian random variable $b$ and an approximately flat distributed one, $x$; as a consequence, the PDF of the first term 
is given by an exponential integral function \begin{eqnarray}
f(h)=\dfrac{1}{c} \operatorname{Ei}\left(-\frac{h^2}{8\pi \sigma_{\Lambda}}\right)\,,
\end{eqnarray}
where $c$ is a normalization constant, which we further approximate by a Gaussian with zero mean and the same standard deviation as $b$, which is equivalent to replacing $x$ by its mean. In this approximation, the variance of the second term for large $n$ and $D$ is the same as the one of the first term,
\begin{equation}
\sigma_2 \approx \sigma_1 \approx  \frac{9a}{\sqrt{5n}}\,.
\end{equation}

The third term in (\ref{htildedia}) has the largest variance and is thus the most important one.  To compute the variance, we ignore the $-1$ in $(J_k^2-1)$ right from the start, since the corresponding contribution to the sum is negligible in the large $D,\tilde{n}$-limit. The remaining term is proportional to 
\begin{eqnarray}
E_{kk}\equiv \sum_{J_1,\dots, J_D=1}^{\tilde{n}} x J_k^2=N_p\sum_{J_k=1}^{\tilde{n}} x J_k^2\,,
\end{eqnarray} 
whose variance is 
\begin{eqnarray}
\sigma_{E_{kk}}=\sqrt{2C\tilde{n}^{D+1}\left(\frac{\tilde{n}^3}{3}+\frac{\tilde{n}^2}{2}+\frac{\tilde{n}}{6}\right)}
\end{eqnarray}
according to equation (\ref{sigmaEkk}) in App.~\ref{app:a2}. As a consequence, the third term has approximately a normal distribution with variance
\begin{eqnarray}
\sigma_3&=&\mu_\Lambda\frac{9}{{\tilde{n}}^2}\sigma_{E_{kk}}=\sqrt{\frac{2C}{6}\frac{(1+\tilde{n})(1+2\tilde{n})\tilde{n}^2}{\tilde{n}^D}}\frac{9}{\tilde{n}^2}\,,
\label{sigma3}
\end{eqnarray}
where $C\approx 0.11$ from App.~\ref{app:a2} eq.~(\ref{defC}) and $\mu_\Lambda =1/\tilde{n}^D=1/n$. The main variability in the Hessian stems from the trigonometric functions and their random arguments as opposed to the Gaussian pre-factors $\Lambda_{J_1,\dots,J_k}$. Thus, we could have worked with constant pre-factors $\Lambda_{J_1,\dots,J_k}=\mu_\Lambda$ from the start, without loosing any crucial information.

\subsubsection{Off-Diagonal}
The derivation of the variance for off-diagonal elements of the Hessian, $k \neq l $, is analogous to the one for the diagonal ones in the preceding section; the only difference is the appearance of two sums, over $J_k$ and $J_l$, instead of a single sum over $J_{k}$, which changes some numerical factors. For the sake of brevity, we omit the detailed derivation, and note that the variance of the three terms appearing in
\begin{equation}
\tilde{H}_{kl}=\frac{9}{n^2} \sum_{\tilde{N}_p} \bigg(-\sum^n_{J_k=1}\sum^n_{J_l=1}\Lambda_{J_1,\dots,J_D}(J_kJ_l-2)+\sum^n_{J_k=1}\sum^n_{J_l=1}bx(J_kJ_l-1) +\sum^n_{J_k=1}\sum^n_{J_l=1}\mu_{\Lambda}x(J_kJ_l-1)\bigg).
\label{htildeoffdia}
\end{equation}
become 
\begin{eqnarray}
\tilde{\sigma}_1 &\approx&  \tilde{\sigma}_2 \approx \frac{3a}{\sqrt{n}}\,,\\
\tilde{\sigma}_3 &=& \sqrt{\frac{2C}{12}\frac{(1+\tilde{n})(1+2\tilde{n})(\tilde{n}^2+\tilde{n})}{\tilde{n}^D}}\frac{9}{\tilde{n}^2}\,,
\label{sigmatilde3}
\end{eqnarray}
where we used (\ref{varianceEkl}) for the computation of $\tilde{\sigma}_3$. The sum over $\tilde{N_p}$ in (\ref{htildeoffdia}) is shorthand for summing over $J_1,\dots, J_D$ without $J_k,J_l$.

\subsubsection{Summary \label{sec:summaryhessian}}
The Hessian at a critical point of the axionic potential in (\ref{defpotential}) is comprised of a deterministic part, set by the height of the potential $V_{\text{c}}$ at said point, and a random, approximately Gaussian component $\tilde{H}_{kl}$,
\begin{eqnarray} 
{H}_{kl}=-\frac{9}{n^2}V_{\text{c}}+\tilde{H}_{kl}\,,
\end{eqnarray}
for $k,l=1,\dots,D$. $H_{kl}$ is therefore, to a good approximation, a Gaussian random variable with mean
\begin{eqnarray}
\mu_{kl} \approx \dfrac{9}{\tilde{n}^2}(1-V_{\text{c}})
\end{eqnarray}
 and variance
\begin{equation}
 \sigma_{kl}  = \begin{cases} \sigma_{\text{dia}} \approx \sqrt{\dfrac{3}{2}}\sigma_3 \approx \dfrac{3.98}{\sqrt{n}} &\mbox{for $k=l$} \\
\sigma_{\text{offdia}} \approx \sqrt{\dfrac{3}{2}}\tilde{\sigma}_3 \approx \dfrac{2.2}{\sqrt{n}} &\mbox{for $k \neq l$}\,,\end{cases} 
\label{sigmadiaandoff}
\end{equation}
see (\ref{sigma3}) and (\ref{sigmatilde3}). We used $a\ll 1$ above to argue that the variance is set be $\sigma_3$ and $\tilde{\sigma_3}$. For the original potential in  (\ref{defpotential}), $\Lambda_{J_1,...J_D}$ is a flat distributed random variable on $[0,1]$. We recomputed the variance for $\sigma_2$ and find
\begin{eqnarray} 
\sigma_1^2&=&\frac{1}{2}\sigma_3^2 \,, \\
\tilde{\sigma}_1^2&=&\frac{1}{2}\tilde{\sigma}_3^2 \,,
\end{eqnarray}
based on results from App.~\ref{sec:distribution}. 
Thus, (\ref{sigmadiaandoff}) should provide a decent approximation for the variance of the Hessian's elements for the potential (\ref{defpotential}) as well.

\begin{figure}[tb]
    \centering
        \subfloat[Diagonal \label{fig:comparehessiandirectdia}] {\includegraphics[width=0.47\textwidth]{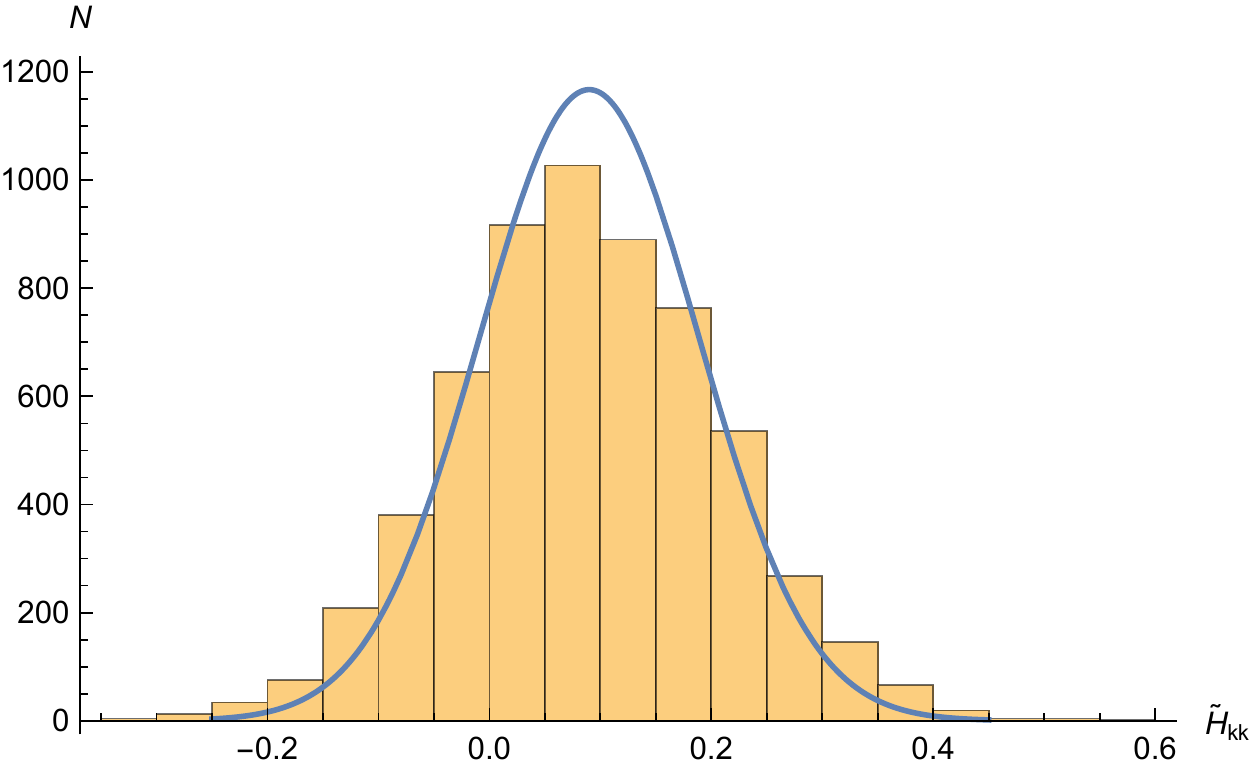}}
        \qquad %
        \subfloat[Off-Diagonal \label{fig:comparehessiandirectoffdia}] {\includegraphics[width=0.47\textwidth]{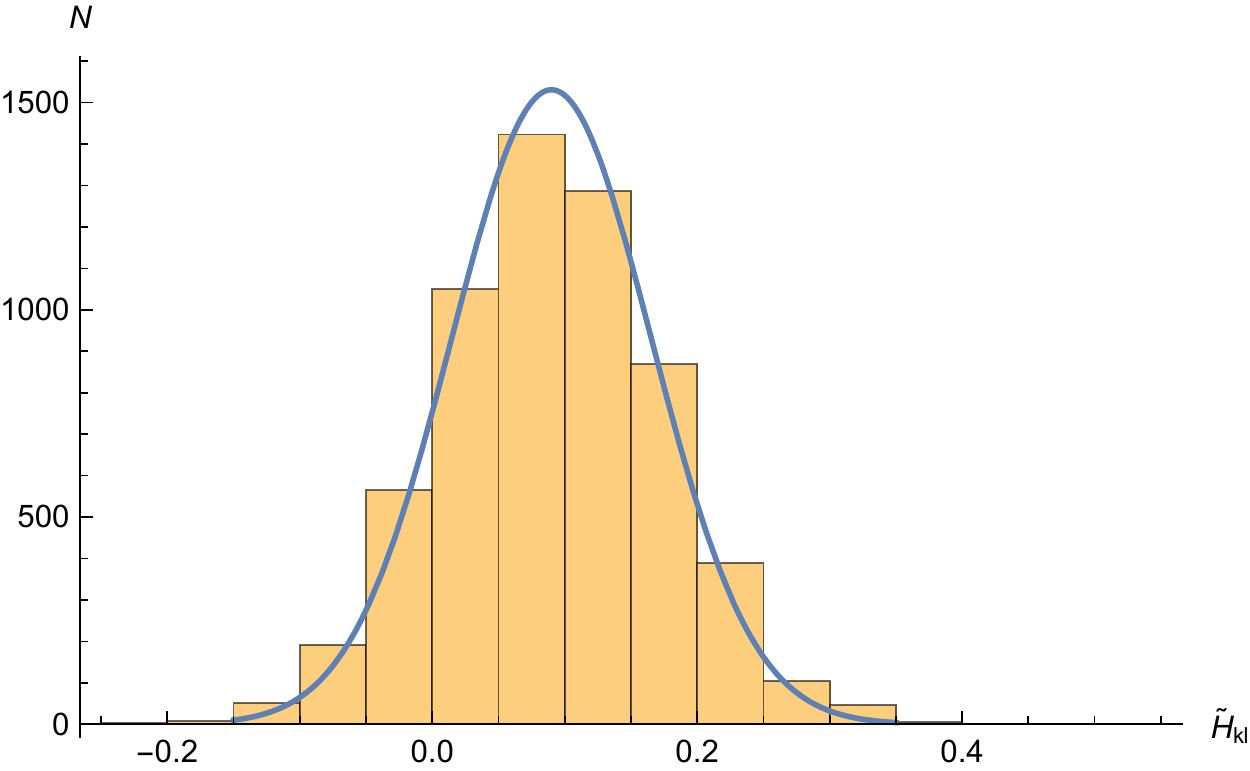}}
\caption{In panel (a) we compare the diagonal entries $\tilde{H}_{kk}$, obtained directly by computing Hessian of the potential ($2000$ realizations, $\tilde{n}=10$, $D=3$) at random points and adding  $9V_{\text{c}}/\tilde{n}^2$, with the theoretically derived Gaussian with mean $\mu_{\text{dia}}=9/\tilde{n}^2$ and variance $\sqrt{3/2}\sigma_{3}$ in (\ref{sigma3}). Panel (b) shows a comparison of the off-diagonal entries of the Hessian at random points, after adding $9V_{\text{c}}/ \tilde{n}^2$ to it, with the theoretically derived Gaussian with mean $\mu_{\text{dia}}=9/\tilde{n}^2$ and variance $\sqrt{3/2}\tilde{\sigma}_{3}$ in (\ref{sigmatilde3}).}
 \label{fig:comparehessiandirect}
\end{figure}

The histogram in Fig.~\ref{fig:comparehessiandirect} shows an exemplary comparison of numerical results for the distribution of entries of Hessians with our analytic approximation: we compute the Hessian of the potential at random points and 
add $9V_{\text{c}}/\tilde{n}^2$ to it, where $V_{\text{c}}$ is the height of the potential at this point. We chose $\tilde{n}=10$ and $D=3$. The solid line is our analytic approximation, not a fit.
We see that the diagonal entries and off-diagonal entries of the Hessian are to a good approximation normally distributed with mean $\mu_{\text{dia}}=(1-V_{\text{c}})9/\tilde{n}^2$ and variances $\sigma_{\text{dia}}$ and $\sigma_{\text{offdia}}$ respectively, see (\ref{sigmadiaandoff}). 

\section{Dynamics on Multi-field Axionic Potentials\label{sec:inflation}
}
In this section, we wish to investigate questions pertaining to the dynamical evolution of axions, particularly where they end up and whether or not inflation is likely to take place. 

\subsection{ The distribution of Minima based on Random Matrix Theory \label{subsec:whyrmtfails}}
Given the Hessian derived in Sec.~\ref{subsec:Hessian}, we can compute the probability that a given critical point $\phi_{\text{c}}$ at height $V_{\text{c}}$ is a minimum by means of random matrix theory in the large $D$-limit, see App.~\ref{sec:rmt} equation (\ref{Pmin}). Given the mean and variance of the Hessian's elements, this probability becomes
\begin{eqnarray}
 P_{\text{min}}(V_{\text{c}})=\frac{1}{2}\bigg(1+\erf\bigg(\frac{9 (1-V_{\text{c}})}{2\sigma_{\text{offdia}}\tilde{n}^2}\bigg)\bigg) \cdot \exp\left(-\dfrac{\ln(3)}{4}(D-1)^2\right)\,,
\label{Pminhessian}
\end{eqnarray}
with $\sigma_{\text{offdia}}$ in (\ref{sigmadiaandoff}). Note that $P_{\text{min}}$ depends on $\tilde{n}$ and $D$ separately and not on $n=\tilde{n}^D$. However, since $\tilde{n}$ is no summation index any more, one can use non-integer values of $\tilde{n}$ according to any desired choice of $n$ and $D$. Evidently, the probability for a random critical point to be minimum becomes exceedingly suppressed in the large $D$ limit. Furthermore, for large $\tilde{n}$ and $D$, that is $\sigma_{kl}\ll 1$, $P_{\text{min}}$ approaches the universal value, see (\ref{Pmin2}),
\begin{equation}
P_{\text{min}} \approx \begin{cases} \exp\bigg(-\dfrac{\ln(3)}{4}(D-1)^2\bigg) &\mbox{for $V_{\text{c}} <1$ ,} \\
0 & \mbox{for $V_{\text{c}} > 1$ .}\end{cases}
\label{Pmin2main}
\end{equation}

Let's consider the large $D$ limit and ask, what is the probability that no minimum is found when starting from $V_{\text{ini}}>1$ until some $0<V<1$? We first recall that  the average potential difference between neighbouring critical points is
\begin{eqnarray}
\overline{\Delta V}\approx \frac{1}{\sqrt{\beta n}}
\end{eqnarray}
according to (\ref{deltav}), with $\beta \approx 1.9$. Therefore, on the descent from $V_{\text{ini}}$ down to $V$, one encounters on average $n_c=(1-V)/\overline{\Delta V}\approx (1-V)\sqrt{\beta n}$ critical points that have a non-zero probability of being a minimum. Thus, the probability of not finding a minimum is
\begin{eqnarray}
P_{\text{nomin}}(V)&\approx& \left(1-P_{\text{min}}\right)^{n_c} \label{approxpnomin}\\
&\approx& 1-\sqrt{\beta n}(1-V)\exp\left(-\frac{\ln(3)}{4}(D-1)^2\right) \,,
\end{eqnarray}
where we kept the leading order term in $P_{\text{min}}\ll 1$ in the last step. (\ref{approxpnomin}) is a monotonic function, which explains correctly that minima are less likely to be encountered as $D$ is increased. However, its derivative with respect to $V$, which should yield a distribution similar to those depicted in Fig.~\ref{fig:VminoverD}, does not have a peak and is therefore insufficient to explain the histograms. One cause of this discrepancy is the omission of dynamics: if fields evolve on the potential, or one simply follows the gradient, minima act as attractors to trajectories in their vicinity. Thus, it it is not surprising that more minima are found in our numerical studies than the naive application of random matrix theory and counting of critical points would have one believe. Another shortcoming becomes evident in the limit $D\rightarrow \infty$, while keeping $\tilde{n}=\text{const}$, yielding
\begin{eqnarray}
\lim_{D,n\rightarrow \infty} P_{\text{nomin}}(0)=1\,.
\end{eqnarray} 
In this limit, runs that encounter a minimum appear to be of zero measure, even though the potential is sharply bounded from below at $V=0$ so that every trajectory must find a minimum. Again, this discrepancy is easily understood: nowhere in the derivation of $P_{\text{min}}$ or $\overline{\Delta V}$ did we impose the lower bound of  the potential. This discrepancy should become important once the lower boundary is approached, that is once $V\sim \overline{\Delta V}$.

In the above argument, we assumed that we start each run with $V_{\text{ini}}>1$. Of course, one can marginalize over the initial values, as we did in the computation of the histograms in Fig.~\ref{fig:VminoverD}, which leads to the same qualitative results:
under the assumption of a flat prior for $V_{\text{ini}}$ over the range $[0,2]$, we can integrate over $V_{\text{ini}}$ to obtain the probability that no minimum is reached up to a final value of $V$,
\begin{eqnarray}
\nonumber P_{\text{nomin}}(V)&=&\frac{1}{2}\big(1-P_{\text{min}}(V)\big)^{\frac{1-V}{\overline{\Delta V}}}\cdot \Big[1+\frac{\overline{\Delta V}}{\ln \big(1-P_{\text{min}}(V)\big)\big(1-V\big)}\Big]-\frac{\overline{\Delta V}}{2\ln \big(1-P_{\text{min}}(V)\big)\big(1-V\big)}\,.\\
&& \label{Pnomin}
\end{eqnarray}
$P_{\text{nomin}}$ is a again a monotonic decreasing function with $P_{\text{nomin}}(0)\neq 0$ that approaches  the universal value
$P_{\text{nomin}}(0)\approx 1$ in the large $D$-limit, showing the same shortcomings as (\ref{approxpnomin}).

\begin{figure}[tb]
\begin{center}
 {\includegraphics[width=0.55\textwidth]{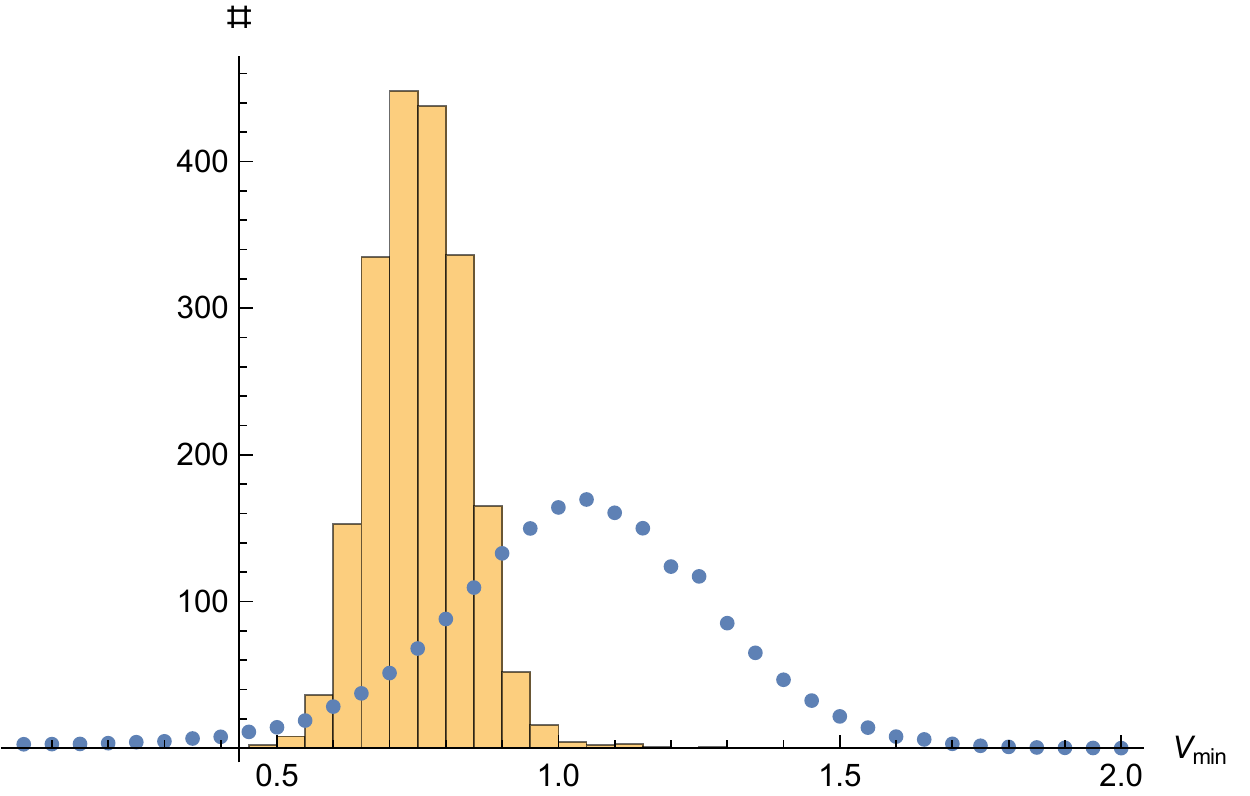}}
\caption{{The height of the dynamically reached minima $V_{\text{min}}$ for $D=3$ and $n=125$ (yellow histogram, as in Fig.~\ref{fig:Vminhist:a}), with the theoretical result computed by taking the derivative of $1-P_{\text{nomin}}$ in (\ref{Pnomin2}) with respect to $V$  numerically and renormalizing the result accordingly (blue dots). While some aspects are properly recovered, such as the presence of a peak and the order of magnitude for the position of the peak and the variance, strong discrepancies are present. Furthermore, these discrepancies are not alleviated by increasing $D$, see discussion in the text.  }}
  \label{fig:CompareRMTnumerics}
  \end{center}
\end{figure}

 For small $D$ and $\tilde{n}$, (\ref{Pmin2main}), and thus (\ref{Pnomin}), is not a good approximation.  One can use (\ref{Pminhessian}) to get an expression for $P_{\text{nomin}}(V)$ that can be evaluated numerically,
\begin{eqnarray}
 P_{\text{nomin}}(V)=\int^2_{V} \prod^{n_c}_{j=0} (1-P_{\text{min}}(V_{\text{ini}}-j\cdot \Delta V)) \frac{1}{V_{\text{ini}}-V} dV_{\text{ini}}\,,
 \label{Pnomin2}
\end{eqnarray}
However, this expression fails as well to explain the observed histograms quantitatively, see Fig.~\ref{fig:CompareRMTnumerics} for an exemplary comparison for $D=3$ and $n=125$: taking a derivative with respect to $V$ yields indeed the expected peak, but its position and width show large discrepancies in comparison to the actual histograms. 

As noted in \cite{Marsh:2013qca}, a further conceptual shortcoming of applying random matrix theory is that the diagonal elements of the Hessian are not independent from each other or the off-diagonal elements, and vice versa. This dependence becomes unimportant in the large $D$ limit, essentially due to the central limit theorem, but it is present for low $D$, as we were able to observe in numerical simulations: the Hessians at random points in our axionic landscape is indeed distinguishable from random matrixes in the GOE for low $D$ and $\tilde{n}$.

To conclude, the probability that a critical point is a minimum as derived in (\ref{Pminhessian}), which is at the heart of the above expressions, does not yield a quantitatively satisfactory explanation of the observed histrograms in Fig.~\ref{fig:Vminhist} for any value of $D$. We attribute this discrepancy to three primary reasons,
\begin{itemize}
\item the omission of dynamical effects (the primary reason),
\item the omission of constraints, such as the potential's lower bound (important for $V\sim \overline{\Delta V}$),
\item the omission of correlations between the Hessian's elements (relevant for low $D$ and $\tilde{n}$).
\end{itemize}
Evidently, $P_{\text{min}}$ in (\ref{Pminhessian}) does not coincide with the empirical result of dividing the average potential difference between encountered critical points $\overline {\Delta V}$ by the mean potential difference to the next minimum, $d V$, 
\begin{eqnarray}
P_{\text{min}}  \neq \frac{\overline{\Delta V}}{dV}\,.
\end{eqnarray} 
To make quantitative progress, let us define the function
\begin{eqnarray}
R(D) \equiv  \dfrac{dV}{ \overline{\Delta V}}\,,
\label{defRfunction}
\end{eqnarray}
for which we are able to provide an empirical analytic approximation in the next section.

\subsection{Empirical Distribution of Minima \label{Sec:empirical}}

\begin{figure}[tb]
\begin{center}
  {\includegraphics[width=0.55\textwidth]{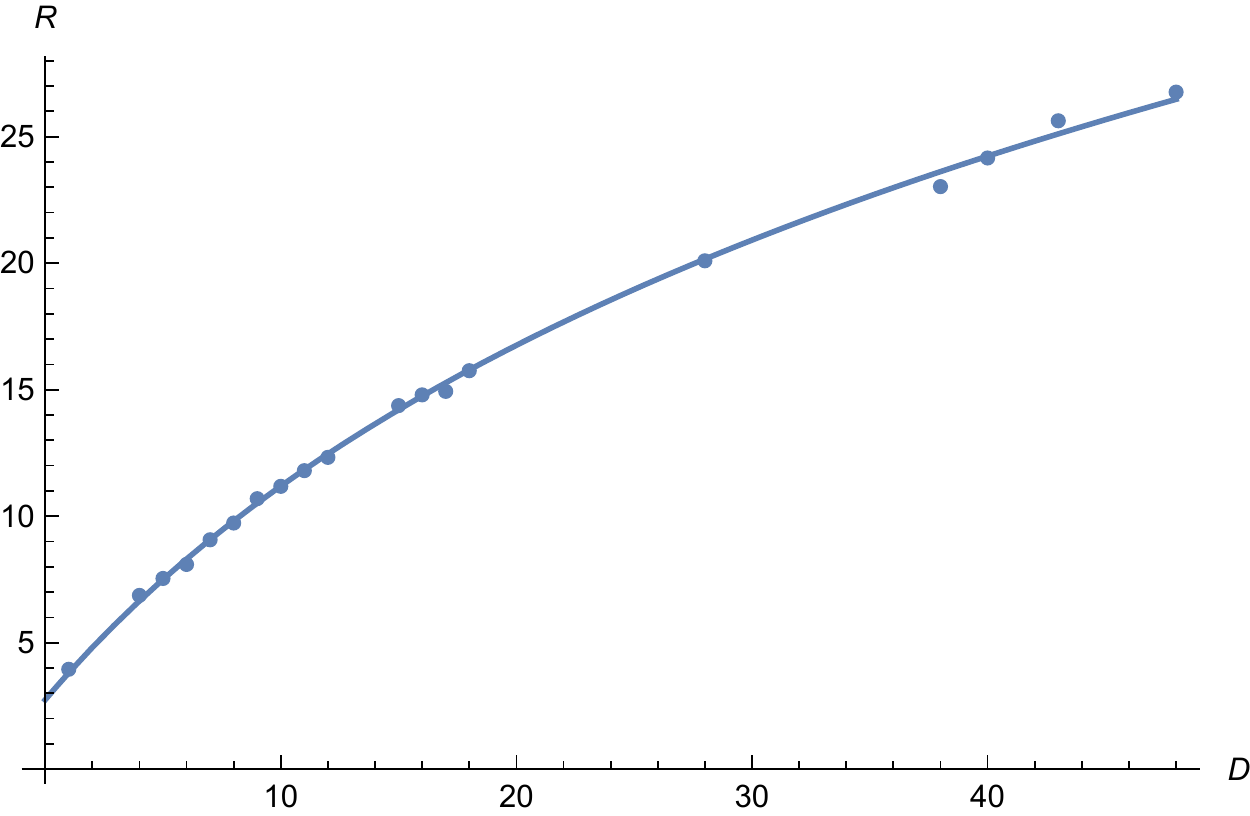}}
\caption{{$R(D)$ in (\ref{defRfunction}) is computed numerically for selected $D$ and $n=125$ by dividing the distance to the nearest minimum by $\overline{\Delta V}$ in (\ref{deltav}). Each blue dot is the average over $200$ random points. The solid line is the Ansatz in (\ref{AnsatzR(D)}) for $R(D)$ with $a= 16.38$ and  $b= -41.56$. Choosing a different $n\gg D$, such as $n=500$, $n=1500$ or $n=5000$, does not change the result significantly.  \label{fig:R(D)}}}
\end{center}
\end{figure}

To model $R(D)$ defined in (\ref{defRfunction}) we make the Ansatz
\begin{eqnarray}
R(D)=a \ln(15+D)+b \label{AnsatzR(D)}\,.
\end{eqnarray}
Comparison to numerical results for the potential in (\ref{defpotential}) yield the parameters ($1\sigma$ errors)
\begin{eqnarray}
a&\approx& 16.38\pm 0.14\,,\\
b&\approx& -41.56 \pm 0.48\,,
\end{eqnarray}
as a the result of a least mean square fit, see Fig.~\ref{fig:R(D)} for details. We confirmed numerically that $R(D)$ does not depend significantly on $n$.
(\ref{defRfunction}) together with (\ref{deltav}) yields
\begin{eqnarray}
 dV(D,n)&=&R(D)\cdot \overline{\Delta V} \, \\
   &=&\left(a\ln(15+D)+b\right) \frac{\Gamma \left( \frac{D+1}{2} \right)}{\Gamma \left( \frac{D}{2} \right)} \frac{\sqrt{2}}{\beta \sqrt{Dn}}\label{difftomin}\\
&\approx& \left(a\ln(15+D)+b\right)\frac{1}{\sqrt{n\beta}}   
   \,.  
\end{eqnarray}
for the potential difference to the next minimum, where we used the large $D$ limit in (\ref{deltavlarged}) in the last step. Thus, the expected height of the final resting place becomes
\begin{eqnarray}
V_{\text{min}}(D,n)=V_{\text{ini}}-dV(D,n)\,.
\label{defvmin}
\end{eqnarray}

\begin{figure}[tb]
    \centering
        \subfloat[$\bar{V}_{\text{min}}$ \label{fig:VmincompareoverD:a}] {\includegraphics[width=0.47\textwidth]{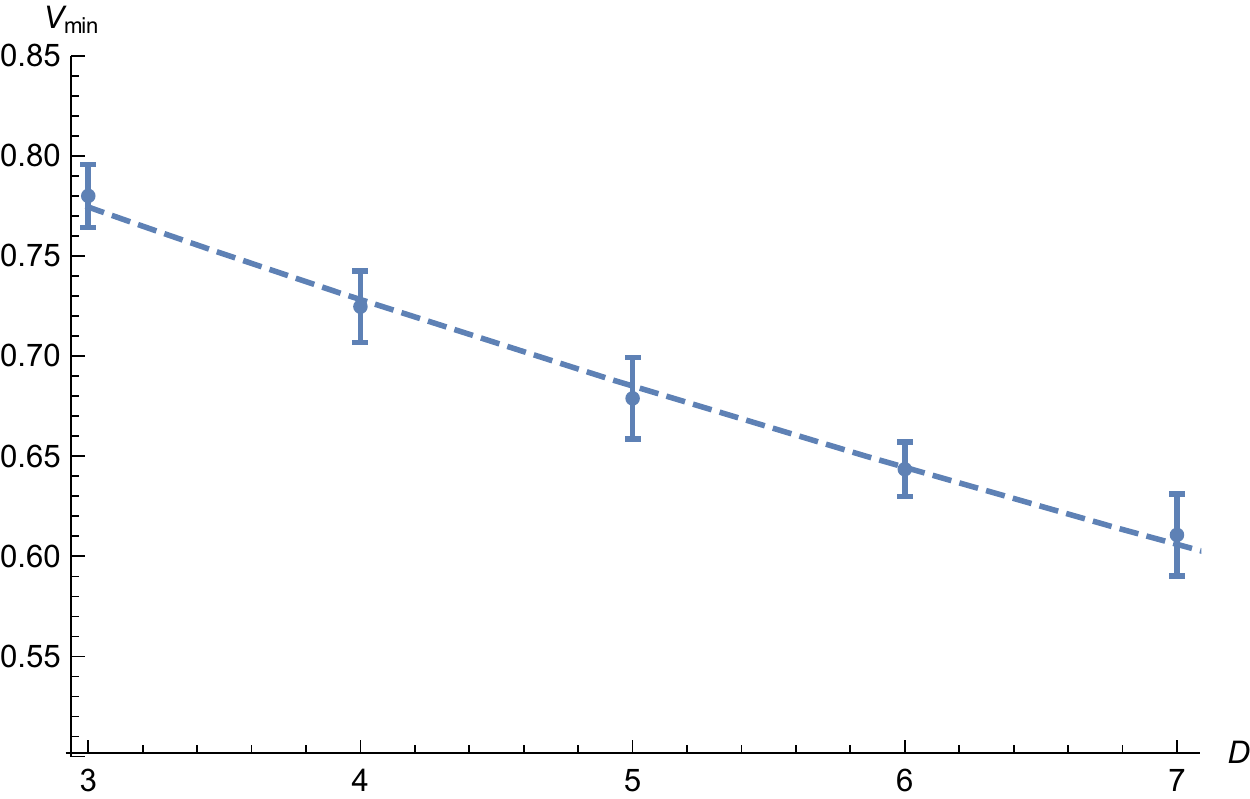}}
        \qquad %
        \subfloat[$\ln{\bar{V}_{\text{min}}}$ \label{fig:VmincompareoverD:b}] {\includegraphics[width=0.47\textwidth]{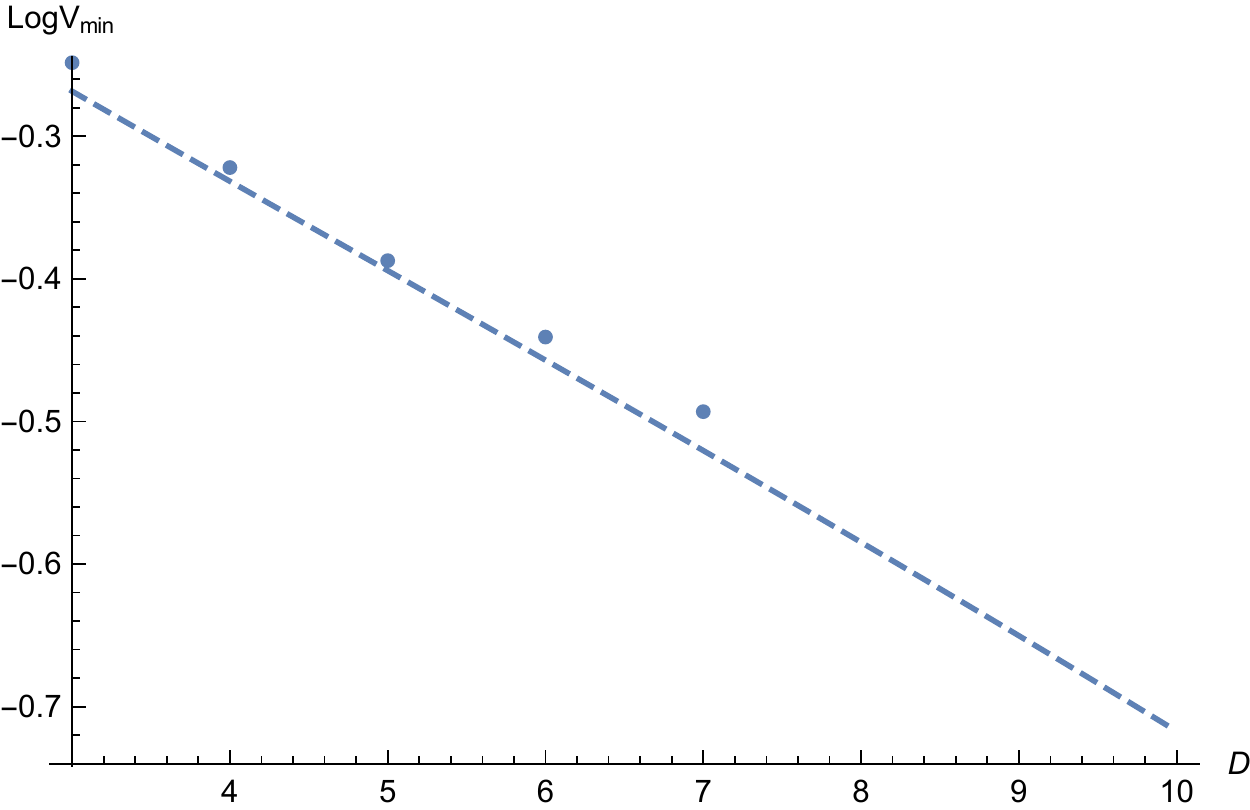}}
\caption{{Comparison of $V_{\text{min}}$ as a function of $D$ given in (\ref{defvmin}) (dashed line, $\tilde{n}^D=n=125$) with the numerical results (average over $30$ realizations), see Fig.~\ref{fig:Vminhist} and Fig.~\ref{fig:VminoverD}. Error bars in panel (a) indicate the $1\sigma$ variance.}}
  \label{fig:VmincompareoverD}
\end{figure}

\begin{figure}[tb]
    \centering
        \subfloat[$\bar{V}_{\text{min}}$ \label{fig:VminoverDlarge:a}] {\includegraphics[width=0.47\textwidth]{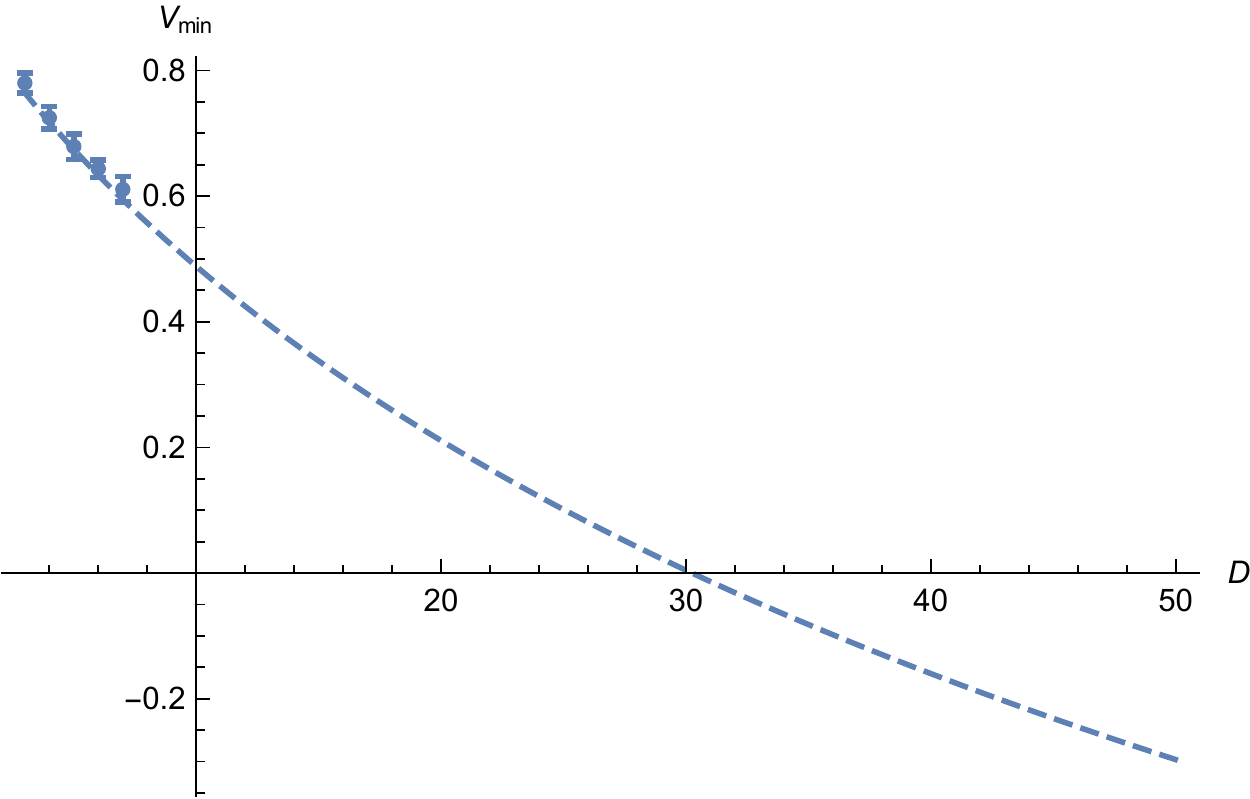}}
        \qquad %
        \subfloat[$\bar{V}_{\text{min}}$ \label{fig:VminoverDlarge:b}] {\includegraphics[width=0.47\textwidth]{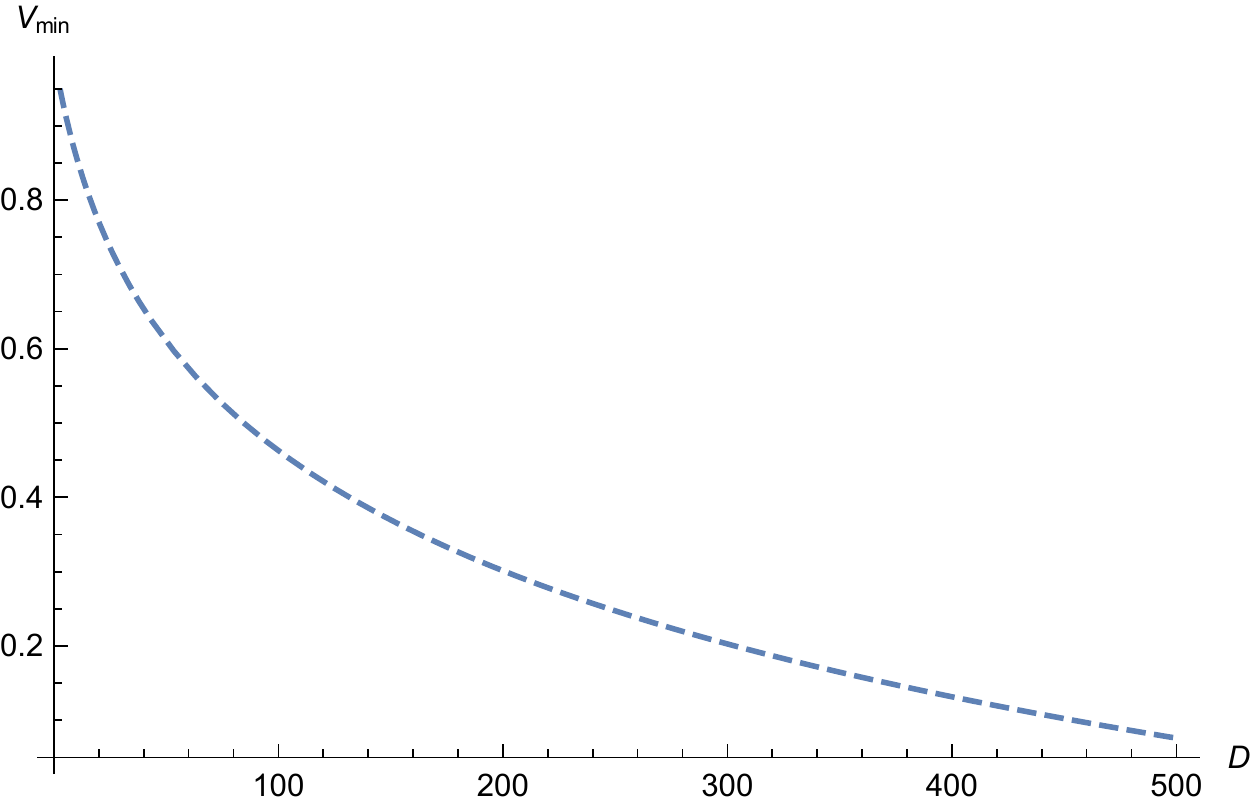}}
    \caption{{Panel (a):  $V_{\text{min}}$ over $D$ for $n=125$, as in Fig.~\ref{fig:VmincompareoverD} but for a larger range. (b) $V_{\text{min}}$ over $D$ for $n=1250$. }}
  \label{fig:VminoverDlarge}
\end{figure}

One could obtain a distribution for the final resting place by integrating (\ref{defvmin}) over all possible $V_{\text{ini}}$. Based on Appendix \ref{sec:distribution}, we know that $V_{\text{ini}} \in \mathcal{N}(1,a/\sqrt{n})$, so that the resulting variance is strongly suppressed in the large $n$-limit. Thus, taking $V_{\text{ini}}=1$ should provide a good approximation for large $n,D$. This is in line with our prior theoretical result in (\ref{Pmin2main}) that $P_{\text{min}}=0$ 
if a critical point is at a height $V>1$ in the same limit. However, for low $D$, we observe that some minima are reached at $V>1$, see Fig.~\ref{fig:Vminhist}. Evidently, there is a some dependence of the effective $V_{\text{ini}}$ on $D$. Treating $V_{\text{ini}}$ as a free parameter that is fitted to the data in Fig.~\ref{fig:VmincompareoverD:a} for low $D$ yields
\begin{eqnarray}
V_{\text{ini}}=1.037 \pm 0.0023 \,,
\end{eqnarray}
close to our theoretical expectation. Given this value of $V_{\text{ini}}$,  we compare (\ref{defvmin}) with the position of the histograms' peaks (see  Fig.~\ref{fig:VminoverD})   in Fig.~\ref{fig:VmincompareoverD} and Fig.~\ref{fig:VminoverDlarge}. We observe good agreement of (\ref{defvmin}) with the numerical results. We would like to highlight that while the five data-points could be fitted by other functions as well, as we did in (\ref{fitlogV}), there is no such freedom left once the results depicted in Fig.~\ref{fig:R(D)} are taken into account, which have a significantly larger reach in $D$.  

The expression for $V_{\text{min}} (D,n)$ in (\ref{difftomin}) can in principle become negative in the large $D$-limit for the numerical values for $a$ and $b$, but it remains positive as long as $n\gg D$ is properly enforced. To be concrete, solving $V_{\text{min}}=0$ for $D$ yields $D=\exp((\sqrt{n\beta}-b)/a)$. Setting this expression equal to $n$ and solving for $n$ gives $n\approx 5075$, that is, above this value one would need $D>n$ to get $V_{\text{min}}=0$. The largest ratio of $n/D$ for which $V_{\text{min}}=0$ is possible is $n/D\approx 6$, which occurs at $n\approx 565$. Thus, as long as $D<n/6$, $V_{\text{min}}$ remains positive for all combinations of $n$ and $D$, consistent with the  definition of the positive semi-definite potential in (\ref{defpotential}). As such, our result is not applicable to the cases studies in \cite{Higaki:2014mwa} due to the low values of $n$ employed there, see for instance the $n=13$ and $D=8$ case that we reproduced in Fig.~\ref{fig:Vmin813}.

\subsection{Discussion: Implication for Vacuum Selection and the Cosmological Constant Problem}\label{disc:ccproblem}

In line with the results in \cite{Battefeld:2012qx}, we found, by direct application of random matrix theory in the large $D$ limit, that almost all critical points are indeed saddle points: the probability for a minimum scales as $\propto \exp(-D^2\ln(3)/4)$, see eqn.~(\ref{Pmin2main}). However, this probability can not be used directly to identify the likely resting place after dynamical evolution on the landscape, since minima act as attractors to neighbouring trajectories, see the discussion at the end of Sec.~\ref{subsec:whyrmtfails}. Thus, directly searching for all minima does not necessary shed light on our vacuum.

Based on an empirical Ansatz for the average potential difference to the nearest minimum in (\ref{AnsatzR(D)}), we were able to predict the most likely height of the dynamically reached minimum in (\ref{defvmin}), which scales as
\begin{eqnarray}
V_{\text{min}}(D,n)\propto \left(1-\text{const.}\,\frac{\ln(15+D)}{\sqrt{n}}\right)\,. \label{finalVmin}
\end{eqnarray}
This logarithmic dependence on $D$ is in stark contrast to the exponential dependence one might naively expect based on (\ref{Pmin2main}). The latter was used in \cite{Battefeld:2012qx} to argue that exceedingly low lying minima would be reached dynamically as the dimensionality of field space is increased. There, only limited numerical tests for low $D$ were available, which were consistent with such an assertion. However, as we have seen here, the preference of low lying minima is considerably weaker than anticipated.

An important consequence pertains to the cosmological constant problem: even with fine tuning of $n$ and $D$, the height of the most likely final resting place can usually not be brought anywhere near the observed value of the cosmological constant. Hence, even though for large $n$ and $D$ an axionic landscape permits a dense distribution of vacua such that some ought to be at the correct height close to the observed value of the cosmological constant \cite{Bachlechner:2015gwa}, it is extremely unlikely that any of them would be reached dynamically via classical evolution, since they are too close to the lower boundary of the potential at $V=0$. One could add a finely tuned negative constant to the potential such the $V_{\text{min}}$ is at the correct height, but this would simply shift the fine tuning problem to another constant.

We did not account for tunnelling, which may very well reduce the lifetime of vacua at intermediate height drastically, see \cite{Tye:2007ja,Sarangi:2007jb,Podolsky:2007vg,Podolsky:2008du,Brown:2010bc,Brown:2010mg,Brown:2010mf,Greene:2013ida}. However, based on the observation of fluctuations in the cosmic microwave background radiation, and the absence of observable spatial curvature, we know that no tunnelling took place during the last sixty e-folds of inflation. Since any prior tunnelling event needs to leave the axions sufficiently high in the potential to allow for inflation and reheating above the energy scale needed for nucleosynthesis, we can conclude that the subsequent, classical evolution leads to the problem elaborated on above.

To conclude, we find that the cosmological constant problem is not alleviated at all by the mere presence of many fields and vacua, leaving yet again anthropic arguments in the framework of eternal inflation as a last resort \cite{Weinberg:1987dv,Bousso:2008bu,Polchinski:2006gy,Bousso:2012dk}. We expect similar results to hold for other multi-field potentials, as long as they are sufficiently random.

\subsection{Inflation on Axionic Potentials}\label{inflation}
We wish to investigate dynamics, particularily the feasibility of inflation, on axionic potentials in the limit $n\gg D\gg 1$. Thus, we need to solve the Friedmann equation in a flat universe
\begin{eqnarray}
3H^2=V(\phi_1,\dots, \phi_D)+\sum_{k=1}^D\dot{\phi}_k^2\,,
\end{eqnarray}
in conjunction with the generalized Klein-Gordon equations 
\begin{eqnarray}
\ddot{\phi}_k+3H\dot{\phi}_k^2=-\frac{\partial V}{\partial \phi_k}\equiv -V_k\,.
\end{eqnarray}
To achieve slow roll inflation for an extended period of time, the potential slow roll parameters, i.e. the ratio between the first and second derivative of the potential along the trajectory to the potential, need to be small.

We first note that the potential at a random point is approximately normal distributed with mean $\bar{V}=1$ and variance $\sigma_V \propto 1/\sqrt{n}$, see eqn.~(\ref{varianceV}). Thus, as we increase $n$, the potential on the whole becomes flattened around the mean, which is a simple consequence of keeping the potential  normalized so that its values stay in the interval between zero and two. 
Next, consider the slope in the steepest direction, as measured by 
\begin{eqnarray}
\epsilon =\frac{1}{2}\sum_{k=1}^{D} \left(\frac{V_k}{V}\right)^2\,.
\end{eqnarray} 
Since $V_k$ is approximately Gaussian distributed with zero mean and $\sigma_{V_k}\approx 1/\sqrt{n}$, see (\ref{sigmaVk}), we showed that the expectation value for the slope in (\ref{defvprime}) has a $\chi$-distribution with mean
\begin{eqnarray}
\overline{V^{\prime}}
\approx \sqrt{D/n}\,,
\end{eqnarray} 
see eqn.~(\ref{expectationvalueVprime}).  Approximating $V$ by its mean value, we arrive at the estimate
\begin{eqnarray}
\epsilon \approx{\frac{D}{2n}}\,.
\end{eqnarray}
and for an individual direction, we get $\epsilon_k \sim 1/(2n)$. Thus, picking a point at random, these slow roll parameters become strongly suppressed and inflation becomes likely. This is consistent with the scaling of
the potential difference to the next critical point $\overline{\Delta V} \propto 1/\sqrt{n}$, see (\ref{deltav}): one has to traverse a smaller and smaller potential difference to find the next critical point, i.e. the potential is exceedingly flat and inflation should be likely. In addition, if one chooses the mass matrix in (\ref{massmatrix}) such that the ratio of the smallest to the second smallest eigenvalue is small, the KNP mechanism is likely to yield a flat direction suitable for inflation.

To build more realistic models, one can shift the potential by the expected height of the final resting place, $V_{\text{min}}$ in (\ref{defvmin}), that is $V\rightarrow V-V_{\text{min}}(D,n)$. Half of the trajectories land on average in a final minimum above (but considerably closer) to zero.
These potentials are fine tuned. Such a shift reduces Hubble friction, increases the expected slow roll parameters and consequently reduces the likelihood of inflation to some degree. However, since $1-V_{\text{min}}\sim \mathcal{O}(1)$ for all values of $n,D$, see the discussion at the end of Sec.~\ref{Sec:empirical}, our conclusions with respect to the likelihood of inflation should remain qualitatively valid. 

To test this expectation, we computed numerically the distribution for the number of e-folds
\begin{eqnarray}
\mathcal{N}=\int H \text{d}t
\end{eqnarray}
 for the unshifted and shifted potential $V\rightarrow V-V_{\text{min}}(D,n)$, focussing on trajectories terminating at $V_{\text{min}}>0$. For $D\lesssim 6$, we find $\mathcal{N}\sim \text{few}$ before the trajectories settle into a minimum and eternal inflation results. For larger $D$, numerical methods become increasingly time consuming, since the computational cost scales with the number of terms in the potential. 
  The computational limitation of globally defined random potentials is well known and can be alleviated by constructing the potential locally around the trajectory \cite{Marsh:2013qca,Battefeld:2014qoa,Bachlechner:2014rqa}. We are currently using such locally defined potentials, based on generalized Dyson Brownian potentials \cite{Battefeld:2014qoa}, to model axionic potentials and perform a quantitative analysis of inflation, including the computation of cosmological perturbations as in \cite{Price:2015qqb} via an extension of MultiModeCode \cite{Price:2014xpa}. These results will be presented in a follow-up publication since they go beyond the scope of this article.
  
\section{Conclusions}\label{conclusion}
Motivated by the feasibility of inflation on multi-field axionic landscapes due to the KNP mechanism, we investigated the distribution of minima that are reached dynamically, particularly the dependence on the dimensionality of field space. Our potentials are bounded from below at $V=0$ and rescaled, such that $V\leq 2$ holds. Furthermore, cross couplings of the fields are included. 

In numerical experiments for $D \leq 7$, we found the peak of the distribution to be shifted to lower values as $D$ is increased. To explain its position, we derived the statistical properties of the potential as well as its first and second derivative at well separated, random points in the limit of many fields and sources, $n\gg D \gg 1$. Together with an estimate for the distribution of critical points, we computed the distribution of minima. This analytic result recovers some qualitative aspects of the distribution, but fails at a quantitative level. We attribute this discrepancy to the difference between counting all minima and the methods by which minima are reached dynamically in a cosmological setting. It should be noted that almost all analytic studies in the literature use the former. We conclude that such simplified methods are insufficient for a quantitative assessment of the final resting place after inflation. 

We proceeded by providing a phenomenological expression for the peak of the distribution in (\ref{finalVmin}), which entails a logarithmic dependence on $D$, which does not approach the lower boundary $V=0$ fast. Thus, even in the large $D$ limit, a considerable bare contribution to the cosmological constant is present if it is not cancelled by a fine tuned additive constant. 

We comment briefly on the feasibility of prolonged periods of inflation on such landscapes, which are likely, but postpone a quantitative study to future work, since the computational cost for the globally defined potentials employed in this paper is prohibitive. This problem can be alleviated by modelling the axionic potentials by locally defined ones, as in \cite{Battefeld:2014qoa}. Our preliminary studies are encouraging.

\begin{appendix}

\section{On the Distribution of the Sum of $N$ Non-identically Distributed Uniform Random Variables}
\label{sec:distribution}

\subsection{The Irwin-Hall Distribution}
The Irwin Hall distribution applies to the the sum of $n$ independent and identically uniformly distributed, $U(0, 1)$, random variables.
In the large $n$ limit, it approaches a normal distribution with mean $\mu=n/2$ and variance $\sigma^2=n/12$, see \cite{IrwinHall}.

\subsection{The Distribution of the Sum of $n$ Uniform Random Variabels \label{app:a2}}
We encountered the sum on non-identically, uniformly distributed random variables, such as 
\begin{equation}
E_{kl}=\sum^{n}_{J_1,\dots,J_D=1} 2y_{J_1,\dots,J_D} J_kJ_l\,,
\end{equation}
where $J_j$ runs from 1 to $n$ for each $j$, $j$ runs from 1 to $D$, $y_{J_1,\dots,J_D}=x_{J_1,\dots,J_D}/2$ are independent and identically distributed (i.i.d.) uniform variables on the intervall $(0,1)$ and we dropped a tilde on $n$ to keep the notation simple.  We are interested to derive the distribution of $E_{kl}$.

To this end, consider first the simpler variable
\begin{eqnarray}
a &\equiv& \sum^n_{i=1} i  y_i \equiv \sum^n_{i=1} a_i\,,
 \end{eqnarray}
where $y_i$ are i.i.d. uniform variables on $(0,1)$. Note that $a$ is the sum of $n$ random variables $a_i$, which are random variables on $U(0, i)$, with $i=1,2,\dots,n$. In line with the Irwin Hall distribution, we can approximate the PDF of $a$ by a normal distribution with 
\begin{eqnarray}
\mu_a&=&\frac{1}{2}\sum^n_{i=1}i=\frac{n(n+1)}{4}\,, \\
\sigma_a^2&=&C n \mu_a  \,,
\end{eqnarray}
where $C$ is a constant that we computed numerically to
\begin{eqnarray} 
C \approx 0.11\,. \label{defC}
\end{eqnarray}
To get closer to $E_{kl}$, let us consider 
\begin{eqnarray}
b\equiv \sum^n_{i=1} 2 i y_i\equiv \sum^n_{i=1} b_i
\end{eqnarray}
as the next step. Evidently, $b$ is the sum of $n$ random variables $b_i$, which in turn is a random variable on $U(0, 2i)$ for $i=1,2,\dots,n$.
Thus, for large $n$, the distribution of $b$ approaches again a normal distribution, but with twice the mean and four times the variance if compared to the one for $a$,
\begin{eqnarray}
\mu_{b}&=&2\mu_a=\sum^n_{i=1}i=\frac{n(n+1)}{2}\,, \\
\sigma_{b}^2&=&C^\prime n \mu_{b} \quad \text{with}\quad C'=2C \approx 0.22\,.
\label{a2}
\end{eqnarray}
Continuing our approach to $E_{kl}$, consider next
\begin{eqnarray}
d\equiv \sum^n_{i=1} 2i^2 \cdot y_{i}\equiv \sum^n_{i=1}d_{i}\,,
\end{eqnarray}
where the i-th summand, $d_i$, is a random variable on $U(0, 2i^2)$, with $i=1,2,\dots,n$. As usual, the distribution of $d$ approaches a normal one, this time with 
\begin{eqnarray}
\mu_{d}&=&\sum^n_{i=1}i^2=\frac{n^3}{3}+\frac{n^2}{2}+\frac{n}{6}\,, \\
\sigma_{d}^2&=&C' n^2 \mu_{d}\,. \label{sigmaEkk}
\end{eqnarray}
As the last intermediate step, consider
 \begin{eqnarray}
 E\equiv \sum^n_{k=1}\sum^n_{i=1} 2 i k y_i= \sum^n_{k=1}kb \equiv \sum^n_{k=1}E_{k}\,,
 \end{eqnarray}
where $E_k = k b$ has a normal distribution with
\begin{eqnarray}
\mu_{E_k}&=&k\mu_{b}=k\sum^n_{i=1}i\,, \\
\sigma_{E_k}^2&=&k^2\sigma_{b}^2 =k C' n \mu_{E_k}
=k^2C'n\sum^n_{i=1}i \,.
\end{eqnarray}
so that
\begin{eqnarray}
\mu_{E}&=&\sum_{k=1}^n k\mu_{b}=\sum_{k=1}^nk\sum^n_{i=1}i=\frac{n^2(n+1)^2}{4}\,, \\
\sigma_{E}^2&=&C'n\bigg(\sum^n_{k=1}k^2\bigg)\bigg(\sum^n_{i=1}i\bigg) 
=C'\frac{n^4+n^3}{12}(1+n)(1+2n)\,.
\end{eqnarray}
Given the distribution of $E$, we see that
\begin{eqnarray}
E_{kl}&=&\sum^n_{J_1,\dots,J_D=1} 2y_{J_1,\dots,J_D} \cdot J_kJ_l
=\sum^n_{\begin{subarray}{c}J_1,\dots,J_D=1\\
    \text{without }J_k,J_l\end{subarray}\,,
}E= n^{D-2}E
\end{eqnarray}
is a Gaussian random variable with 
\begin{eqnarray}
\mu_{E_{kl}}&=&n^{D-2}\mu_e=\dfrac{1}{4}n^D(n+1)^2\,, \label{meanEkl}\\
\sigma_{E_{kl}}^2&=&n^{D-2} \sigma_{E}^2 =C'n^D\frac{n^2+n}{12}(1+n)(1+2n)\,.\label{varianceEkl}
\end{eqnarray}

\section{Random Matrix Theory }
\label{sec:rmt}

Here, we present some known results of random matrix theory pertaining to fluctuation probabilities of eigenvalues for matrices in the Gaussian Orthogonal Ensemble (GOE) with i.i.d. entries and zero mean, see \cite{Marsh:2011aa}.  
The probability that an $N \times N$ real matrix in the GOE is positive-definite is \cite{Aazami:2005jf,Dean06,Dean08,Marsh:2011aa}
\begin{equation}
 P \propto \exp\bigg(-\dfrac{\ln(3)}{4}N^2\bigg)\,,
 \label{positiveprop}
\end{equation}
which is much smaller than one might naively expect based on the $N$ eigenvalues of such a matrix.

Instead of matrices in the GOE with zero mean, we want to study  eigenvalue fluctuation of $N \times N$ random real symmetric matrices, whose entries have non-zero mean and  different variances for diagonal and off-diagonal entries. Firstly, in the large $N$-limit, we can ignore the difference in variance for the diagonal elements, since it does not enter into the fluctuation probability, see e.g. the lecture notes in \cite{Kemp:2013}.

Based on numerically generating $N \times N$, real, symmetric matrices with off-diagonal entries drawn from $\mathcal{N}(\mu,\sigma)$ as well, we find that the probability for the largest eigenvalue to be positive for $\mu>0$ is approximately given by
\begin{equation}
 P(\lambda_1>0)\approx\frac{1}{2}\bigg(1+\erf\left(\frac{|\mu|}{2\sigma}\right)\bigg)\,.
 \label{plargest}
\end{equation}
The distribution of the remaining eigenvalues obeys Wigner's Semicircle Law, so that
\begin{equation}
P(\lambda_2,...,\lambda_N>0)=\exp\big[-\dfrac{\ln(3)}{4}(N-1)^2\big]\,.
\end{equation}
Thus, the probability that a critical point is a minimum is approximately given by\footnote{ We checked numerically that (\ref{B4}) is a good approximation for the probability that a critical point is a minimum. Note that Wigner's Semicircle Law applies to matrices in the GOE with zero mean of their entries, whereas we are dealing with a non-zero mean. The latter is the cause of the error function in (\ref{B4}), which recovers properly the fact that minima are exceeding rare for $\mu<0$.}
\begin{equation}
P_{\text{min}}\approx P(\lambda_1>0)\cdot P(\lambda_2,...,\lambda_N>0)\approx \frac{1}{2}\bigg(1+\erf\bigg(\frac{|\mu|}{2\sigma}\bigg)\bigg)\cdot\exp\bigg(-\dfrac{\ln(3)}{4}(N-1)^2\bigg)\,,\label{B4}
\end{equation}
where we treat the largest eigenvalue as if it were independent from all the other eigenvalues, i.e., we don't require the remaining eigenvalues to be smaller than $\lambda_1$. If $\mu<0$, equation \ref{plargest} describes the probability that the smallest eigenvalue is less than zero so that the probability that a critical point is a minimum becomes
\begin{equation}
P_{\text{min}}\approx P(\lambda_N>0)\cdot P(\lambda_1,...,\lambda_{N-1}>0)\approx \frac{1}{2}\bigg(1-\erf\bigg(\frac{|\mu|}{2\sigma}\bigg)\bigg)\cdot \exp\bigg(-\dfrac{\ln(3)}{4}(N-1)^2\bigg)\,.
\end{equation}
Hence, for arbitrary $\mu$ we get
\begin{equation}
P_{\text{min}}\approx \frac{1}{2}\bigg(1+\erf\bigg(\frac{\mu}{2\sigma}\bigg)\bigg) \cdot \exp\bigg(-\dfrac{\ln(3)}{4}(N-1)^2\bigg).
\label{Pmin}
\end{equation}
 For $\sigma \ll 1$, the error function in (\ref{Pmin}) approaches a step function, yielding
\begin{equation}
P_{\text{min}} \approx \begin{cases} \exp\big[-\dfrac{\ln(3)}{4}(N-1)^2\big] &\mbox{for $\mu >0$ ,} \\
0 & \mbox{for $\mu<0$ .}\end{cases}
\label{Pmin2}
\end{equation}

\end{appendix}

\acknowledgments
We would like to thank J. Niemeyer for discussions and support. G.W. would like to thank Zi Ye from the Technical University Munich for nice discussions on several mathematical problems involved here.

-----------------------------------------------------------------

\end{document}